\documentclass[1p,review]{elsarticle}

\newcommand{\LF}{1}
\newcommand{\MF}{0.6}
\newcommand{\SF}{0.49}
\usepackage{hyperref}
\usepackage{siunitx}
\usepackage{amsmath}
\usepackage{graphicx}
\usepackage{mhchem}
\usepackage{tabularx}
\usepackage{threeparttable}
\usepackage{multirow}
\usepackage{subcaption}
\usepackage{bm}

\journal{Elsevier}
\begin{document}

\begin{frontmatter}
    \title{Performance evaluation of the 8-inch MCP-PMT for Jinping Neutrino Experiment}

    \author[a,b,c]{Aiqiang Zhang}%

    \author[a,b,c]{Benda Xu\corref{cor1}}%
    \ead{orv@tsinghua.edu.cn}
    \author[a,b,c]{Jun Weng}
    \author[a,b,c]{Huiyou Chen}
    \author[a,b,c]{Wenhui Shao}
    \author[a,b,c]{Tong Xu}
    \author[a,b,c]{Ling Ren}
    \author[d]{Sen Qian}
    \author[a,b,c]{Zhe Wang}
    \author[a,b,c]{Shaomin Chen}

    \cortext[cor1]{Corresponding author}
    \affiliation[a]{
        organization={Department of Engineering Physics},
        addressline={Tsinghua University}, 
        city={Beijing},
        postcode={100084},
        country={China}}
    \affiliation[b]{
        organization={Center for High Energy Physics},
        addressline={Tsinghua University}, 
        city={Beijing},
        postcode={100084},
        country={China}}
    \affiliation[c]{
        organization={Key Laboratory of Particle \& Radiation Imaging (Tsinghua University)},
        addressline={Ministry of Education},
        country={China}}
    \affiliation[d]{
        organization={Institute of High Energy Physics},
        addressline={Chinese Academy of Sciences},
        city={Beijing},
        postcode={100049},
        country={China}
    }
    \begin{abstract}
    Jinping Neutrino Experiment plans to deploy a new type of 8-inch MCP-PMT with high photon detection efficiency for MeV-scale neutrino measurements. This work studies the performance of the MCP-PMTs, including the photon detection efficiency, the charge resolution of the single photoelectron, the transition time spread, single photoelectron response, rates of dark counts and after pulses. We find a long tail in the charge distribution, and combined with the high photon detection efficiency, the overall energy resolution sees substantial improvements. Those results will be provided as the inputs to detector simulation and design. Our results show that the new PMT satisfies all the requirements of the Jinping Neutrino Experiment.

\end{abstract}


\end{frontmatter}
\section{Introduction}
The Jinping Neutrino Experiment~(JNE) under construction is a hundred-ton liquid scintillator detector with Cherenkov and scintillation light readout
 at CJPL II~\cite{li_second-phase_2015,cheng_china_2017} with \SI{2400}{m} rock overburden, targeting solar, terrestrial and supernovae neutrinos~\cite{LetterJNE2017,xu_jinping_2020,xu_innovations_2022,xu_design_2022}.
Photomultiplier tubes~(PMT)~\cite{HAMAMATSUManual} are commonly used to detect individual photons in water Cherenkov~\cite{SNO,SuperK} and liquid scintillator detectors~\cite{KamLAND,JUNO:2015zny}. It converts a photon into a photoelectron~(PE) and then to a measurable electric signal.  For the 8-inch form factor, instead of conventional dynodes, the micro-channel plate~(MCP) PMT multiplies PEs inside the micro-channels, offering faster time response and high gain in a compact size~\cite{HAMAMATSUManual,WANG2012113,MCP-PMTworkgroup:2021hoy}.

Precise measurement of energy spectra demands affordable PMTs to achieve good photo-coverage with high photon detection efficiency~(PDE\footnote{The product of PE collection and quantum efficiencies.}). Cherenkov photons providing a directional measurement of solar neutrinos have \SI{1.5}{ns} timing dispersion at a \SI{10}{m} scale, setting the requirement of time precision to be \SI{\sim 1}{ns}.

The new type of 8-inch MCP-PMT (GDB-6082~\cite{GDB-6082}) is produced by North Night Vision Science \& Technology (Nanjing) Research Institute Co. Ltd. (NNVT). 
Similarly structured 20-inch MCP-PMTs by NNVT were evaluated by the JUNO collaboration to have an average PDE of 28\%~\cite{JUNOMassTesting}.
Characterization of gain, single PE resolution, PDE, transit time spread (TTS) and dark count rate (DCR) is the key step to construct neutrino and dark matter detectors. Such tests have widely been carried out, for 8-inch dynode PMTs (9354KB, R5912, XP1806) at Daya Bay~\cite{DayaBayTesting}, 10-inch dynode PMTs (R7081) at Double Chooz~\cite{DoubleChoozeTesting}, 8-inch dynode PMTs (CR365-02-1) at LHAASO~\cite{LHAASOTesting}, 20-inch dynode PMTs (R12860) and MCP-PMTs (GDB-6203) at HyperKamiokande~\cite{HyperKTesting}, 3-inch dynode PMTs (R12199-02) at KM3NeT~\cite{KM3NetTesting}, 3-inch dynode PMTs (R11410-21) at XENON1T~\cite{XENON1TTesting} and XENONnT~\cite{XENONnTTesting}, and 3-inch PMTs (R12199-01) at IceCube~\cite{IceCubeTesting}.

This work characterizes nine GDB-6082 MCP-PMT samples for the JNE.  The setup of the testing facility is introduced in Section~\ref{SetUp}. The analysis methods and results of gain, charge resolution, PDE, TTS, DCR, shape of single electron response~(SER), pre-pulse and after-pulse are described in Section~\ref{Method}. The boost for energy resolution is discussed in Section~\ref{Result} with a summary in Section~\ref{Summary}.

\section{Experimental setup and procedures}
\label{SetUp}
\begin{figure}[!htbp]
    \centering
    \includegraphics[width=\SF\textwidth]{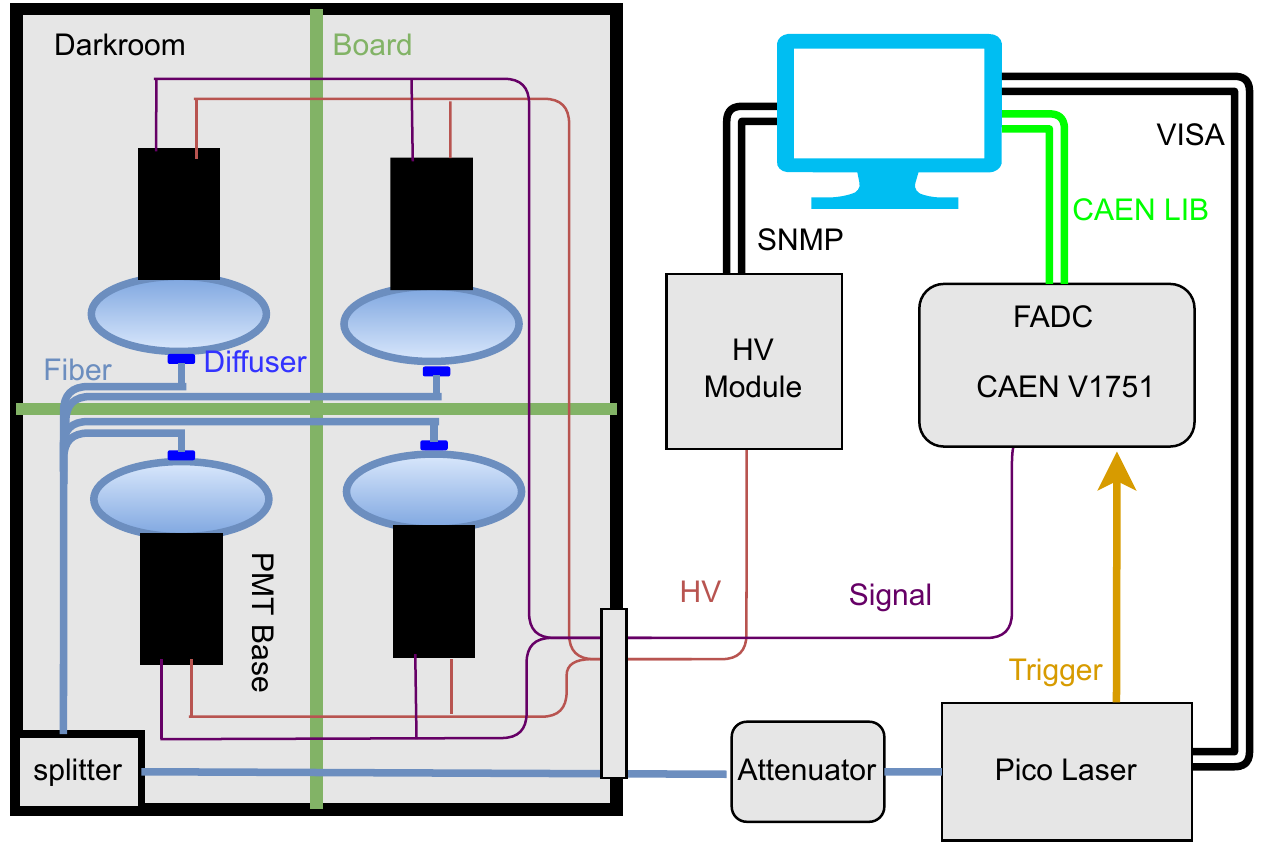}
    \caption{The schematics of the test system. The computer controls the HV module, laser, and FADC directly. The trigger signal from laser and waveforms from PMT are sampled by the FADC. The laser light is split into four channels and diffused after attenuation.}
    \label{fig:facility}
\end{figure}

The schematics of the PMT-test facility is shown in Fig.~\ref{fig:facility}. CAEN V1751 \SI{1}{GS \per s} 10-bit digitizer is used to acquire data~\cite{CAENV1751}. With the dynamic range of 1V, we use the unit of ``1 ADC''~\cite{JUNOPrototype} to represent a quantization level of 1000/1024\,mV in the following sections. Wiener EDS 30330p high voltage (HV) module~\cite{WIENERHV} supplies a positive voltage for each PMT. A picosecond laser~(PiL040XSM) from Advanced Laser Systems~\cite{NTKLaser} produces \SI{34}{ps} light pulses at \SI{405}{nm} to illuminate the PMTs via an attenuator and feeds an electronic trigger signal into the digitizer. The digitizer, the HV module and the laser are controlled by self-developed data acquisition (DAQ) software\footnote{Github repo: \url{https://github.com/greatofdream/CAENReadout}.} based on the CAENDigitizer~\cite{CAENLIB}, Net-SNMP~\cite{SNMP} and PyVISA~\cite{VISA} libraries.

PMTs are installed in dark rooms made of a black light-tight plastic box separated by extruded polystyrene boards into four parts. A fiber splitter distributes attenuated laser light into the four dark rooms. Each channel ends with a \SI{4}{cm\tothe{2}} diffuser plate to spread light onto the top of the PMT photocathode.
A base distributes HV to the pins of each PMT and outputs the amplified pulse from the anode. A CR365 PMT~\cite{BJBS} from Beijing Hamamatsu Photon Techniques Inc. is used as a reference for PDE measurements.  It has a specification of \SI{25}{\percent} quantum efficiency~(QE) at \SI{420}{nm}, and \(0.97 \times \SI{25}{\percent}\) at \SI{405}{nm}~\cite{HAMAMATSUManual}. The PDE at \SI{405}{nm} is estimated to be 17\%, corresponding to a collection efficiency of 0.7--0.8 typical for 8-inch dynode PMTs~\cite{WANG2012113,R5912MOD,RCESpotlight}.

\begin{figure}
    \centering
    \includegraphics[width=0.4\textwidth]{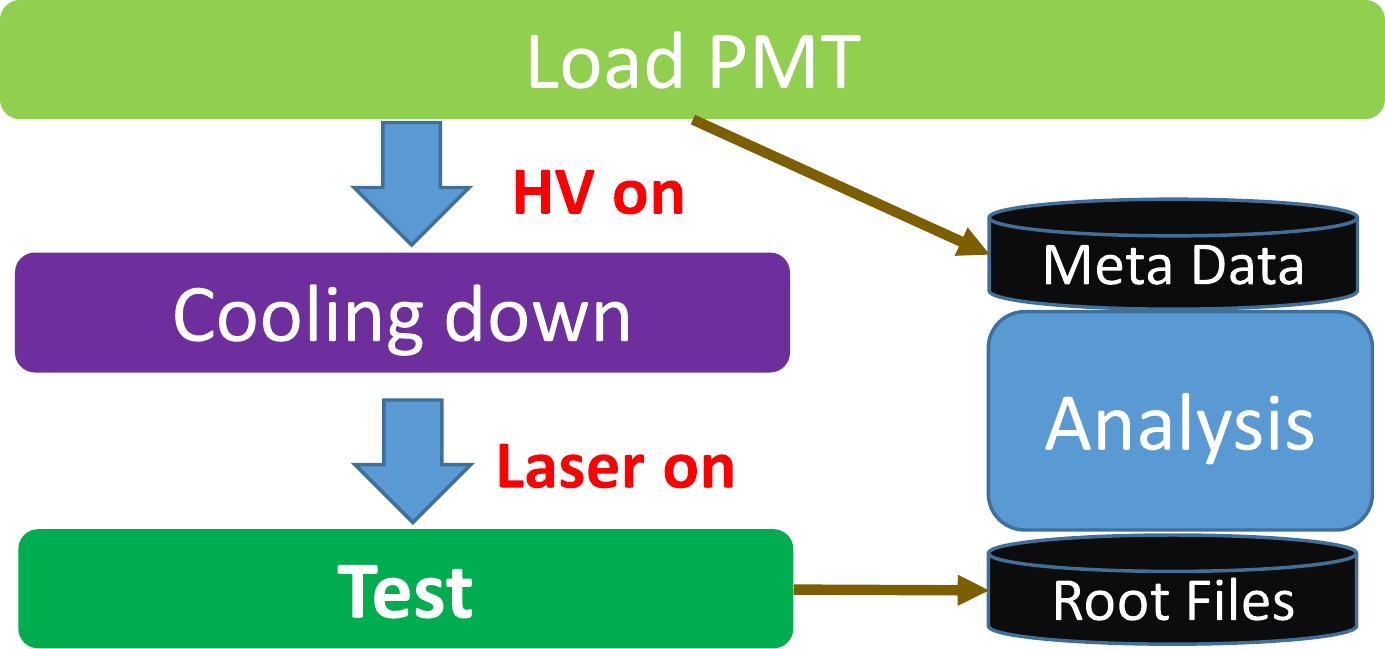}
    \caption{The flow chart of test procedures. PMTs are cooled down before DAQ. The procedures are executed automatically.}
    \label{fig:testingprocedure}
\end{figure}

The test procedures indicated in Fig.~\ref{fig:testingprocedure} are executed by DAQ software automatically. To lower the systematic errors from the light source variation and unknown splitter ratios, we permute the PMTs in the dark rooms to conduct PDE measurements and light source calibration simultaneously in Section~\ref{sec:PDE}. To cool down the DCR, all the PMTs stay in the darkroom with vendor-specified HV for at least 12 hours before the digitizer acquires waveforms with laser on. The waveforms are stored in ROOT~\cite{brun_root_1997} files and analyzed with self-developed software\footnote{Github repo: \url{https://github.com/greatofdream/pmtTest}.}.

\sisetup{separate-uncertainty=true}
\sisetup{multi-part-units=single}
\section{Methods and results}
\label{Method}
\subsection{Preanalysis}
\label{sec:laserstage}

\begin{figure}[!htbp]
    \centering
    \includegraphics[width=\LF\textwidth]{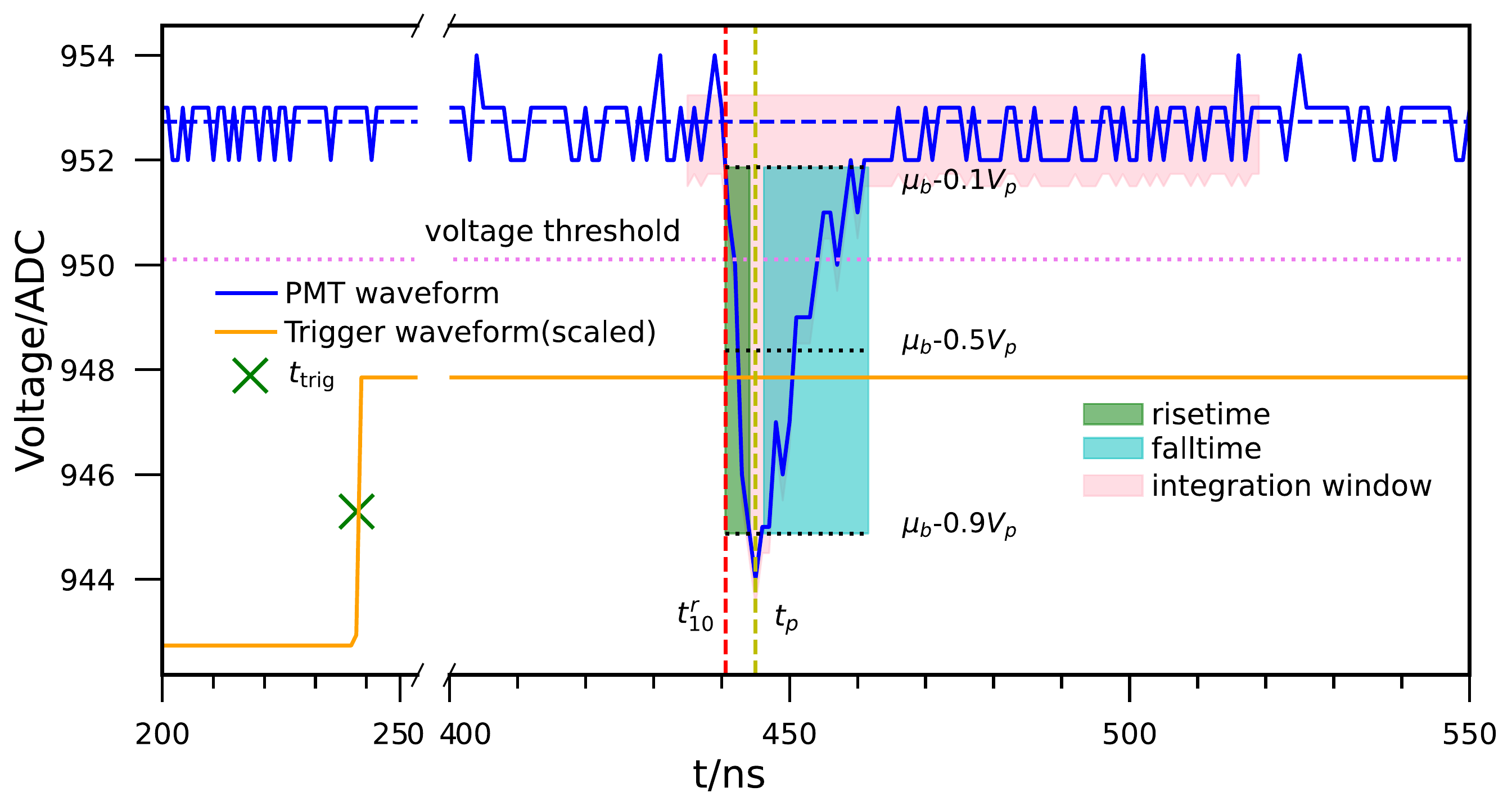}
    \caption{An example of PMT and trigger signals. The orange line is the trigger waveform in arbitrary scale. The green cross is the interpolated trigger time $t_{\mathrm{trig}}$. The solid blue line is the PMT waveform with a PE pulse. The horizontal violet dotted line is the voltage threshold for calculating the baseline shown by the blue horizontal dashed line. The black horizontal dash lines intersect 10\%, 50\% and 90\% of (downward) rising and falling edges. The rise and fall times are the interval lengths between 10\% and 90\% of the respective edges. The red and yellow vertical dashed lines are for the 10\% rising edge and the pulse peak.}
    \label{fig:triggertime}
\end{figure}

The laser intensity is adjusted to the level of $1/20$ occupancy to obtain single PE events. The window size $T_{\mathrm{wave}}$ is \SI{10400}{ns} to include all the possible after-pulses (see Section~\ref{sec:afterpulse}). The rising edge of the trigger waveform from the laser system is linearly interpolated to get the half-height time $t_{\mathrm{trig}}$ at about \SI{250}{ns}, as shown in Fig.~\ref{fig:triggertime}.

In the preanalysis, we select a preliminary window $[t_{\mathrm{trig}},600\,\mathrm{ns}]$ where dark noises (\SI{< 10}{kHz} in Section~\ref{sec:dcr}) and laser pulses are expected to contribute 0.004 and 0.05 counts on average. The peak time $t_p$ is the minimum position in each window, as shown in Fig.~\ref{fig:triggertime}.

The nonzero baseline is estimated from the sidebands $[-200\,\mathrm{ns},-10\,\mathrm{ns}]$ and $[100\,\mathrm{ns},200\,\mathrm{ns}]$ relative to $t_p$. 
To remove potential additional pulses in the sidebands, we define a voltage threshold from a rough estimation of white noise as the horizontal violet dotted line in Fig.~\ref{fig:triggertime} and cut off additional \SI{10}{ns} around each over-threshold time interval. The baseline $\mu_b$ is estimated as the average of the residual sidebands. The peak height $V_p$ of a pulse is the difference between $\mu_b$ and the minimum voltage at \(t_p\).

Over all the waveform samples, a Gaussian $f(t;t_0,\sigma_{t0}^2)$ is fitted to the distribution of $t_p-t_{\mathrm{trig}}$ of pulses whose $V_p$ exceeds \SI{5}{ADC}. We define a new candidate window, about \SI{30}{ns} long, as $[t_0-5\sigma_{t0}, t_0+5\sigma_{t0}]$ to calculate new $t_p$ and $V_p$ by repeating the above procedures.  The contracted time window reduces the dark counts by 10 times, with which we shall conduct the charge and time characterization of the single PE in the following sections.

\subsection{Single-PE charge spectrum and resolution}
\label{sec:noisepeak}

Considering the rise and fall time distributions, the charge $Q$ of a pulse is defined as 
the summation of the baseline-subtracted voltages in a time window $[\SI{-10}{ns}, \SI{75}{ns}]$ relative to $t_p$ as illustrated in the pink region of Fig.~\ref{fig:triggertime}. The input impedance being \SI{50}{\Omega}~\cite{CAENV1751}, the charge in Coulomb is $Q/\SI{50}{\Omega}$.

\begin{figure}[!htbp]
    \centering
    \begin{subfigure}[b]{\SF\textwidth}
        \includegraphics[width=\textwidth]{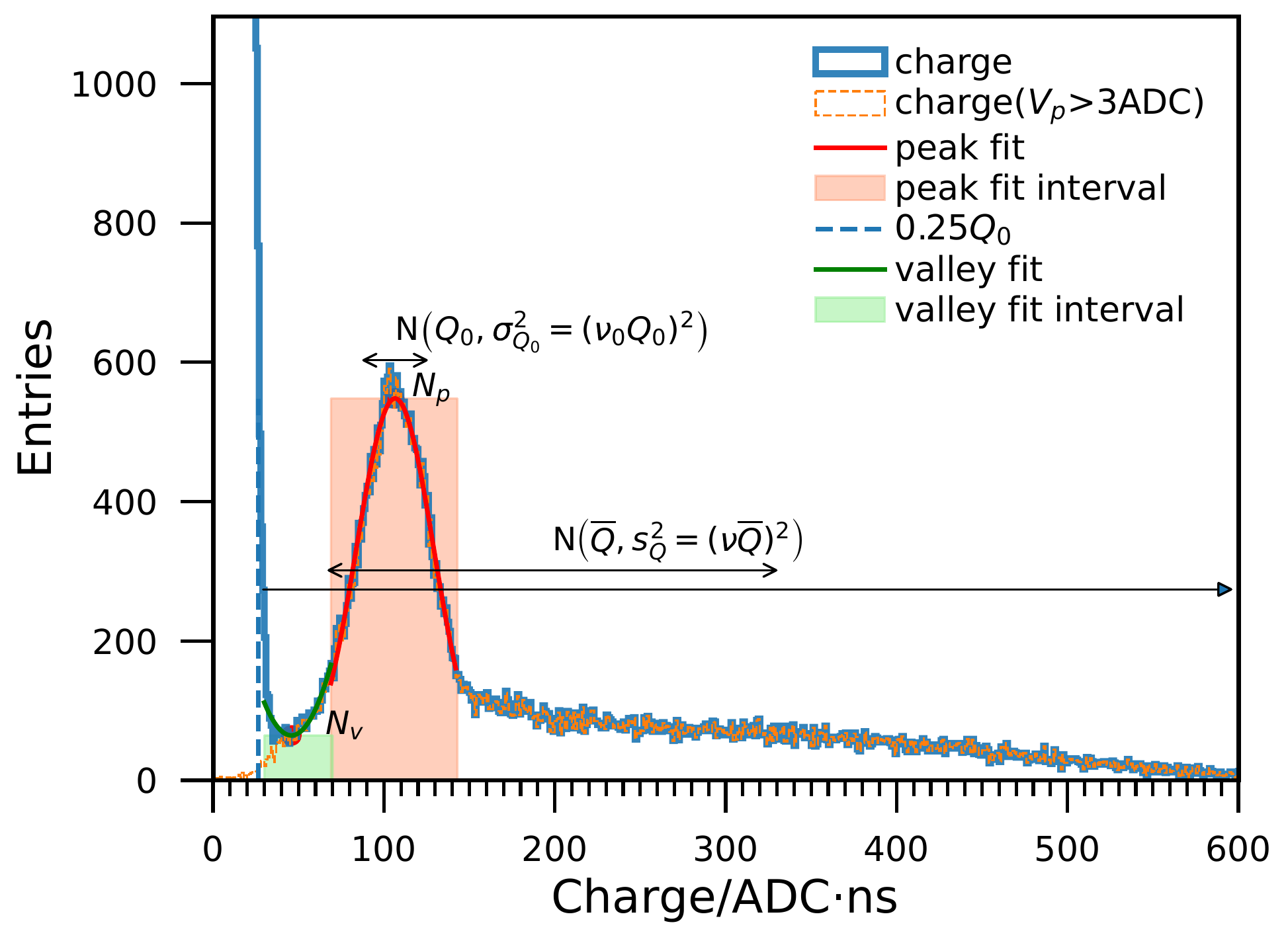}
        \caption{}
        \label{fig:triggercharge}
    \end{subfigure}
    \begin{subfigure}[b]{\SF\textwidth}
        \includegraphics[width=\textwidth]{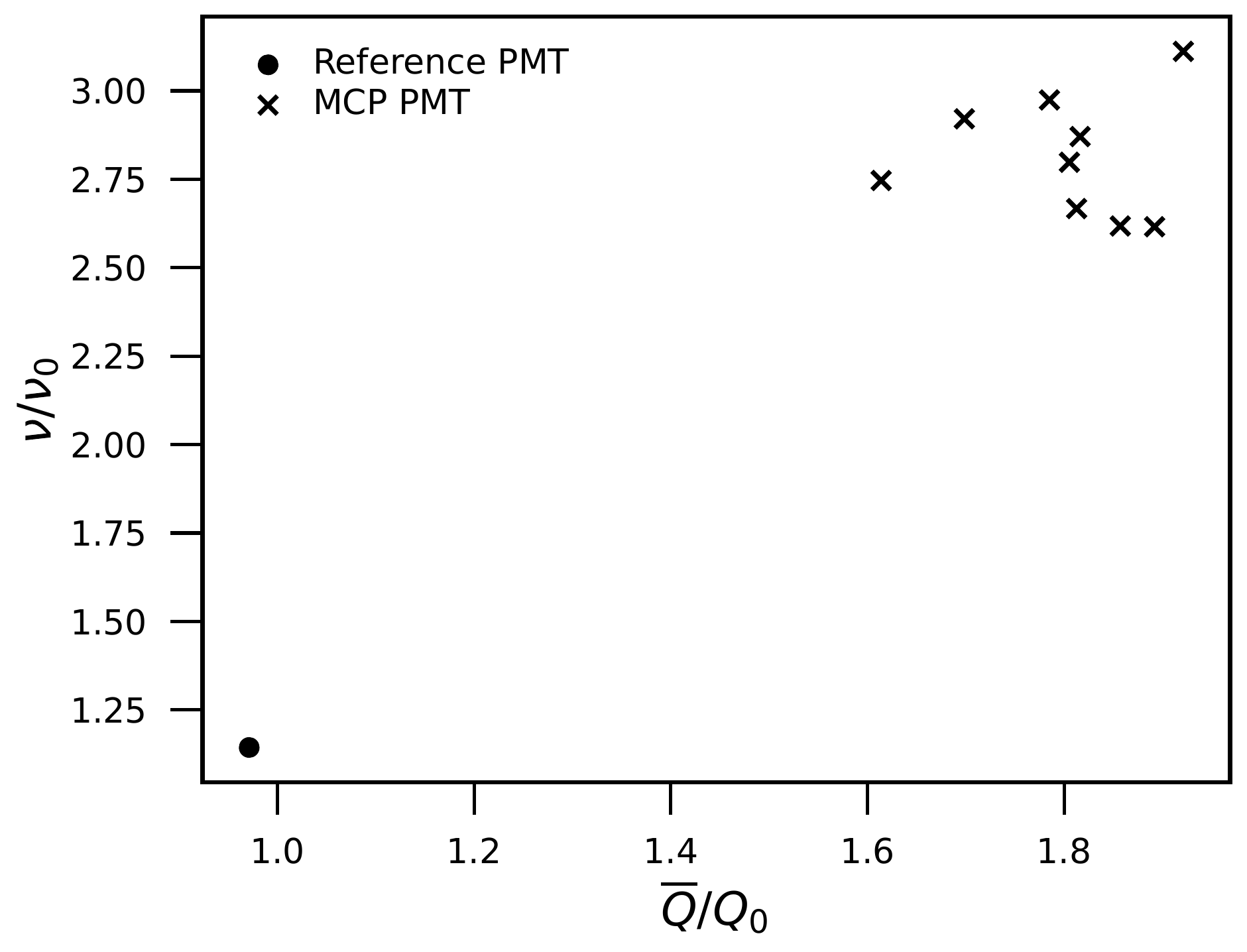}
        \caption{}
        \label{fig:totalchargeCompare}
    \end{subfigure}
    \caption{(\subref{fig:triggercharge}) The long-tailed single-PE charge distribution of an MCP-PMT. The entries around zero are waveforms with no signal. The vertical blue dashed line is the pedestal cut. The orange histogram is the selected waveforms with peak-height cut ($V_p>\SI{3}{ADC}$). The pink and green areas are the fit intervals for the peak and valley parameters. A Gaussian function $f(Q;Q_0,\sigma^2_{Q_0})$ is fitted to the peak of single-PE charge distribution. Mean $\overline{Q}$ and sample variance $s_Q^2$ of selected charge reflect the long-tailed information. (\subref{fig:totalchargeCompare}) The charge and resolution ratios show the effect of the long tail. The point and crosses represent the reference and MCP-PMTs, respectively.
    }
\end{figure}

Such $Q$ in Fig.~\ref{fig:triggercharge} represents the charge of a single PE with negligible multi-PE contributions due to the low occupancy. A long tail is evident in the single-PE charge distribution, which was also found in the NNVT 20-inch MCP-PMTs by the JUNO collaboration~\cite{JUNOMassTesting}. Zhang~et~al.~\cite{JUNOLongtail} proposed a phenomenological parameterization without dedicated consideration of the multiplication process of the PEs. We shall discuss the physical model and solution of the long tail in our future publications.

To describe the peak shape of the $Q$ distribution, a Gaussian function $N^{\mathrm{1e}}f(Q;Q_0,\sigma^2_{Q_0})$ is fitted to the interval $[0.65Q_0, 1.35Q_0]$ via the modified least-square (MLS)~\cite{Cowan1998StatisticalDA} illustrated as the red line in Fig.~\ref{fig:triggercharge}. To remove the influence of the pedestal and describe the long tail, $N^{\mathrm{hit}}$ pulses with $V_p>\SI{3}{ADC}$ and $Q>0.25Q_0$ are selected to calculate the mean $\overline{Q}$ and sample variance $s^2_{Q}$ of $Q$. The Gaussian component ratio $N^{\mathrm{1e}}/N^{\mathrm{hit}}$ ($0.59\pm0.02$ for the nine MCP-PMTs) of charge distribution reflects the significance of long tail.

The gains of the main peak and the entire sample are ${Q_0}/({\SI{50}{\Omega}} e) \approx \num{e7}$ and ${\overline{Q}}/(\SI{50}{\Omega} e)$, $e$ being the charge of an electron. The relative \emph{peak} and \emph{sample resolutions} $\nu_0$ and $\nu$ are defined as ${\sigma_{Q_0}}/{Q_0}$ and \(\left.{\sqrt{s^2_{Q}}}\middle/{\overline{Q}}\right.\). Fig.~\ref{fig:totalchargeCompare} indicates that $\overline{Q}$ is about 1.8 times $Q_0$ for the MCP-PMTs, in agreement with Zhang~et~al.~\cite{JUNOLongtail}. The long tail causes the resolution of MCP-PMTs to degrade from $\nu_0=\num{0.25 \pm 0.02}$ to $\nu=\num{0.69 \pm 0.03}$, which is less pronounced for the reference dynode PMT.

\subsection{Peak-to-valley ratio}
\label{sec:PV}
A parabolic function is fitted to the valley based on MLS in the interval $[-0.15Q_0, 0.25Q_0]$ relative to the least-counted bin of the histogram between the pedestal and the main peak, as shown in Fig.~\ref{fig:triggercharge}. The \emph{valley count} $N_v$ is defined as the minimum of the parabola and \emph{peak count} $N_p$ is the maximum of the Gaussian described in Section~\ref{sec:noisepeak}. The peak-to-valley ratio~(P/V) ${N_p}/{N_v}$ shows the ability to discriminate between electronic noises and a PE signal. The average P/V of MCP-PMTs is about 5.9, significantly higher than that (about 2.4) of the reference PMT.

\subsection{Single electron response}
\label{sec:SER}
We define $t^{10}_r$, $t^{50}_r$, $t^{90}_r$ ($t^{10}_f$, $t^{50}_f$, $t^{90}_f$) as the times of interpolated 10\%, 50\%, and 90\% $V_p$ in the rising (falling) edge as shown in Fig.~\ref{fig:triggertime}.  The rise time $t_r = t^{90}_r - t^{10}_r$, fall time $t_f = t^{10}_f - t^{90}_f$ and full width at half maximum $\mathrm{FWHM} = t^{50}_f - t^{50}_r$ describe the shape of \emph{single electron response}~(SER).  They are measured to be $t_r = \SI{3.72\pm0.15}{ns}$, $t_f = \SI{15.6\pm1.8}{ns}$ and $\mathrm{FWHM} = \SI{9.07\pm0.63}{ns}$ for the nine MCP-PMTs.

To get a smooth SER, signals with $V_p>\SI{3}{ADC}$, $Q \in [0.5Q_0, \SI{1000}{\mathrm{ADC\cdot ns}}]$ and $\mathrm{FWHM} \in [2\,\mathrm{ns}, 15\,\mathrm{ns}]$ are selected to exclude the noise and large pulses. 
An \emph{exGaussian} distribution $f^N(t;\mu_{\mathrm{SER}},\sigma_\mathrm{SER}^2)\otimes f^{\mathrm{Exp}}(t;1/\tau_\mathrm{SER})$~\cite{Luo:2022xrd} is used to fit the SER
, in which $\mu_{\mathrm{SER}}$ is the pulse location, $\sigma_{\mathrm{SER}}$ and $\tau_{\mathrm{SER}}$ model its shape. They are measured to be $\tau_{\mathrm{SER}} = \SI{7.2\pm1.1}{ns}$ and $\sigma_{\mathrm{SER}} = \SI{1.62\pm0.06}{ns}$ for the nine MCP-PMTs.

\subsection{Transit time spread}
\label{sec:TTS}
\begin{figure}[!htbp]
    \centering
    \includegraphics[width=\MF\textwidth]{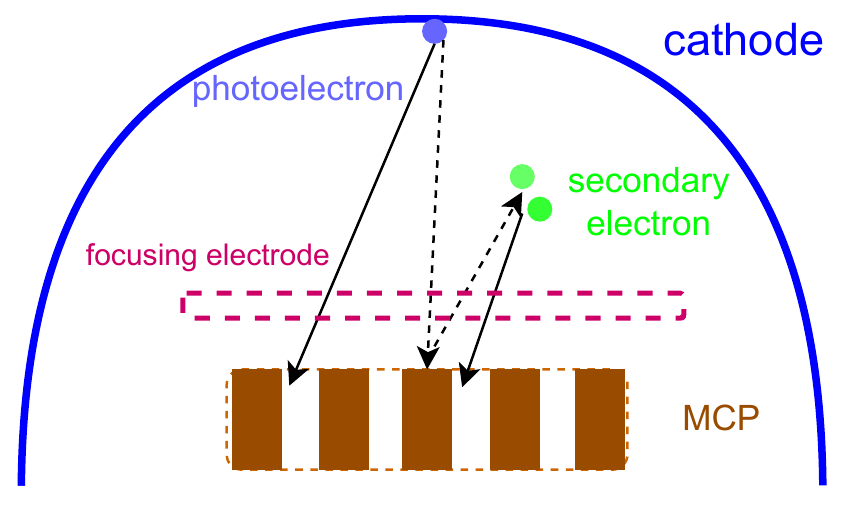}
    \caption{The PEs drift from the photocathode to the MCP and get amplified or scattered.}
    \label{fig:mcpelectron}
\end{figure}

\begin{figure}[!htbp]
    \centering
    \begin{subfigure}[t]{\SF\textwidth}
        \includegraphics[width=\textwidth]{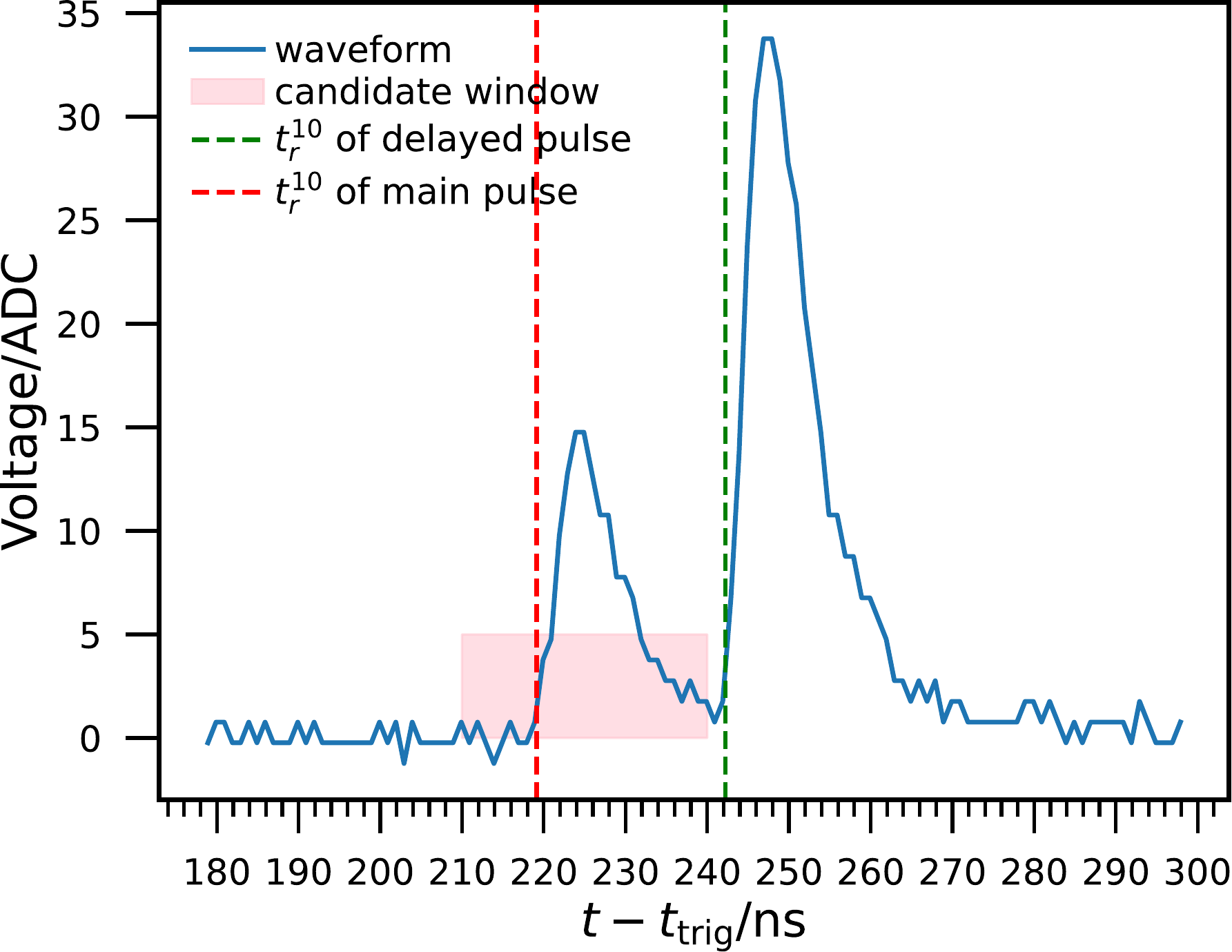}
        \caption{}
        \label{fig:triggerTT2pulse}
    \end{subfigure}
    \begin{subfigure}[t]{\SF\textwidth}
        \includegraphics[width=\textwidth]{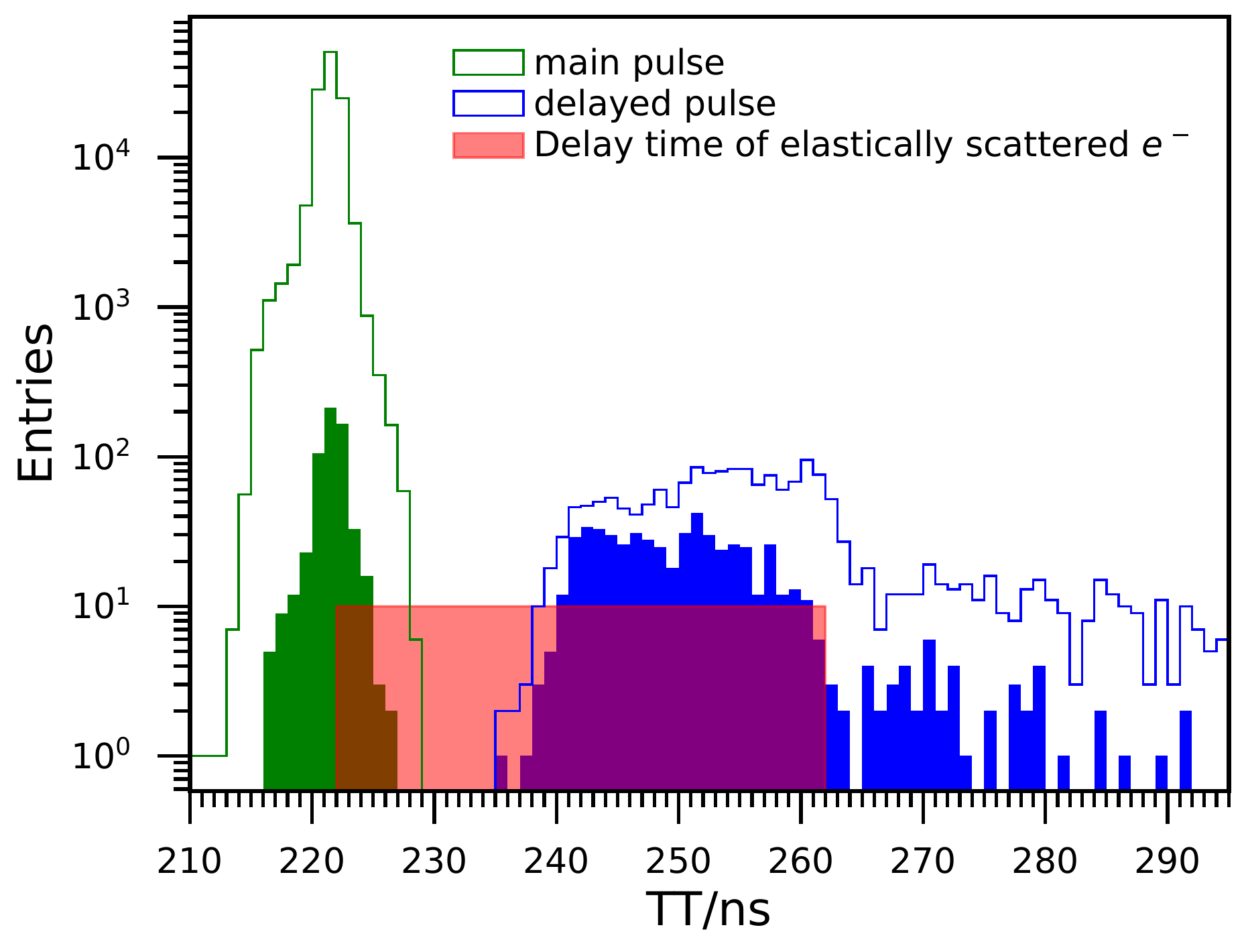}
        \caption{}
        \label{fig:triggerTTlatepulse}
    \end{subfigure}
    \caption{(a) Double-pulse example. The first pulse falls in the candidate window defined in Section~\ref{sec:laserstage}. (b) The green and blue histograms are the main and delayed pulse distributions, respectively. The filled histograms are waveforms containing the main and delayed pulses simultaneously.}
\end{figure}

As shown in Fig.~\ref{fig:mcpelectron}, the PEs from the photocathode drift to the MCP. Using a model of a cathode at \SI{0}{V}, a focusing electrode at \SI{480}{V} and an MCP at \SI{528}{V} to simulate the electric field and the PE trajectory, we find the PEs from the top of the photocathode with \SI{0}{eV} and \SI{3}{eV} kinetic energies have drift times of about \SI{21}{ns} and \SI{18}{ns} respectively. A PE entering an MCP channel is multiplied to be an observable pulse, while that hitting the surface of the MCP gets scattered inelastically into several secondary electrons or elastically into one single electron~\cite{Furman}. The scattered electrons drift in the electric field until finally entering the MCP channels to give delayed pulses~\cite{KM3NetTesting}. Multiple secondary electrons with different kinetic energies may cause two or more pulses with different drift times, one example shown in Fig.~\ref{fig:triggerTT2pulse}.

\begin{figure}[!htbp]
    \centering
    \begin{subfigure}[t]{\SF\textwidth}
        \includegraphics[width=\textwidth]{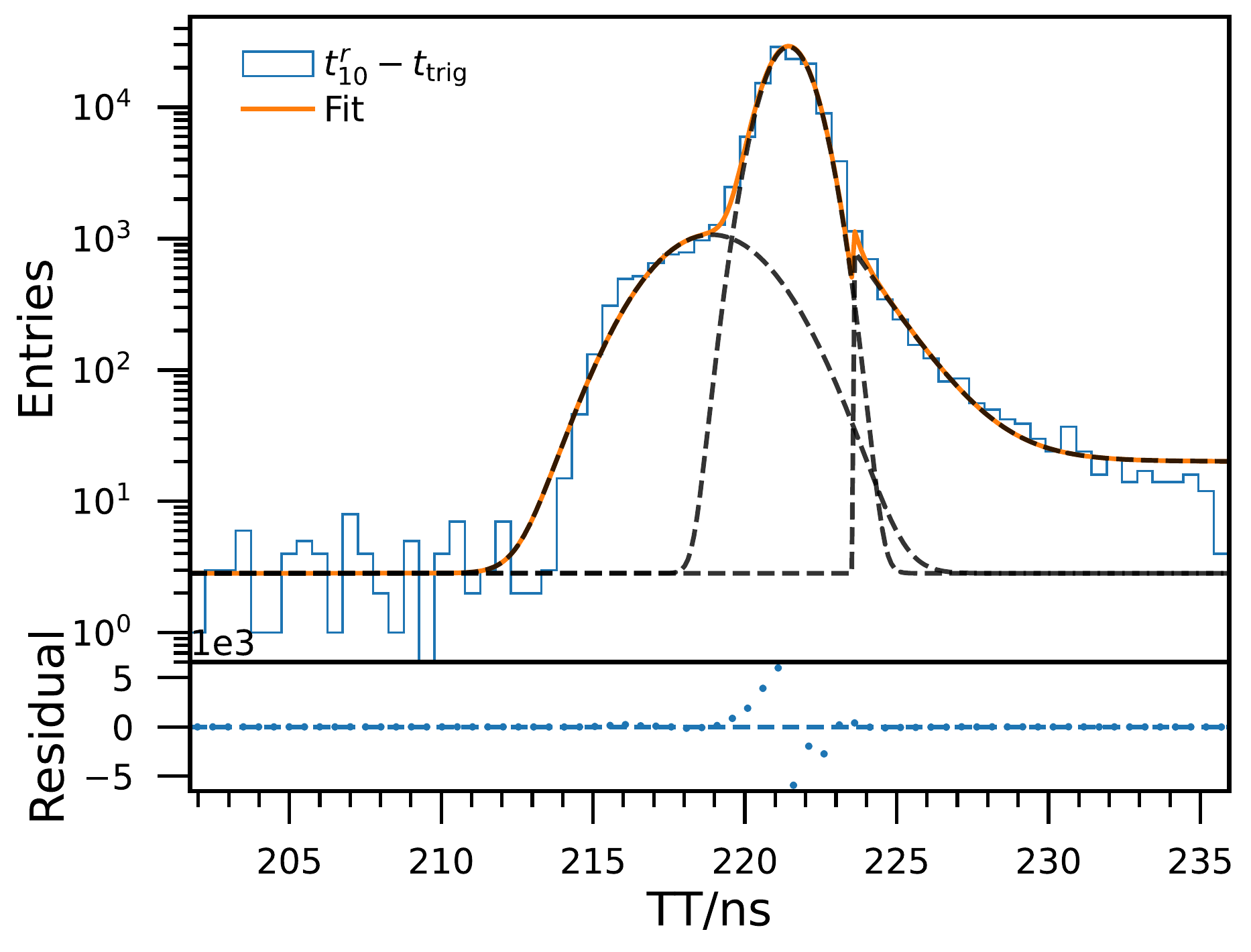}
        \caption{}
        \label{fig:triggerTTSLog}
    \end{subfigure}
    \begin{subfigure}[t]{\SF\textwidth}
        \includegraphics[width=\textwidth]{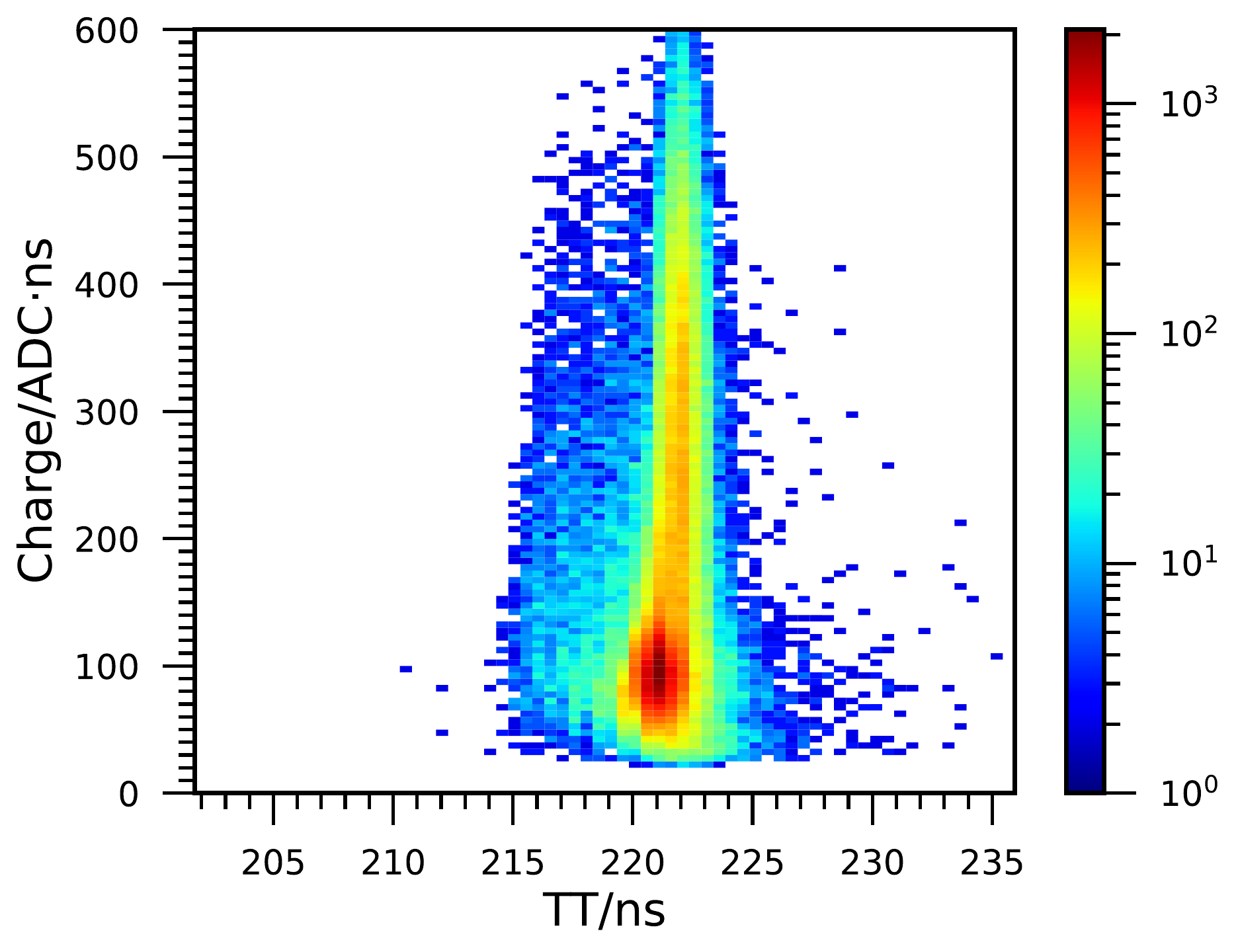}
        \caption{}
        \label{fig:triggerTTS2d}
    \end{subfigure}
    \caption{(a) The top pad is the distribution of TT with the y-axis in the logarithmic scale. The black dashed lines from left to right are early, main and delayed components with a dark noise pedestal. The bottom pad is the residual between data and fit in linear scale. (b) The 2D distribution of TT and charge with the colorbar in logarithmic scale.}
\end{figure}

The time of a PE traveling from the photocathode to the anode is called the \emph{transit time}~(TT). However, the absolute TT is hard to measure. As a convention practically, we refer to the time difference between the trigger signal $t_{\mathrm{trig}}$ and 10\% of the rise time of a PE pulse $t_r^{10}$ as TT instead. The $\mathrm{TT}$ distribution of the MCP-PMTs contains slowly rising and falling edges on both sides of the peak, as shown in Fig.~\ref{fig:triggerTTSLog}. The rising edge is due to the PEs with larger kinetic energies, while the falling one consists of secondary electrons with longer drift times~\cite{longtail}.

Delayed pulses are searched in the interval $[t_0+20\,\mathrm{ns},t_0+80\,\mathrm{ns}]$ to separate them from the main pulses in $[t_0-5\,\mathrm{ns},t_0+5\,\mathrm{ns}]$. The blue histogram in Fig.~\ref{fig:triggerTTlatepulse} is the distribution of the delayed pulses, and the filled one is for those with the main pulses in the same waveform, an example demonstrated in Fig.~\ref{fig:triggerTT2pulse}. The sharp difference between them at about \SI{40}{ns} after the main peak, twice the drift time of PEs from the cathode to the MCP, reasonably illustrates that an elastically scattered electron cannot appear together with a main pulse in a waveform.

In Fig.~\ref{fig:triggerTTSLog}, the main and early components are modeled with Gaussian functions 
$N_1f_1^N(t;\mu_{\mathrm{TT}},\sigma_{\mathrm{TT}}^2)$ and $N_2f_2^N(t;\mu_K,\sigma_K^2)$, with the subscript $K$ standing for high-kinetic-energy PEs. Considering the exponential distribution of kinetic energies of secondary electrons~\cite{Furman,SecondElectron}, $f^\mathrm{Exp}(t;1/\tau_S)$ is suitable to model the delayed component.  We add a constant $b_S$ and a translation of $\mu_{TT} + 3\sigma_{TT}$ to fit the data. The \SI{0.5}{ns}-binned $\mathrm{TT}$ histogram with selection criteria in Section~\ref{sec:noisepeak} is fitted by

\begin{equation}
    \begin{aligned}
        B&+N_1f_1^N(t;\mu_{\mathrm{TT}},\sigma_{\mathrm{TT}}^2)\\
        &+N_2f_2^N(t;\mu_K,\sigma_K^2)\\
        &+b_SH(\mu_{\mathrm{TT}}+3\sigma_{\mathrm{TT}})+N_Sf^{\mathrm{Exp}}\left(t-(\mu_{\mathrm{TT}}+3\sigma_{\mathrm{TT}});\frac{1}{\tau_S}\right)
    \end{aligned}
\end{equation}
in which $H$ is the Heaviside function to restrict the domain of the delayed component, and $B$ is the constant dark noise rate. The early and exponential components are as small as $N_2/N_1 = \num{0.026\pm0.018}$, $N_S/N_1 = \num{0.010\pm0.003}$ and $2b_S\sigma_{\mathrm{TT}}/N_1 = \num{0.00026\pm0.00009}$ for the nine MCP-PMTs.  $\sigma_K$, $\tau_S$ and $\mu_{\mathrm{TT}}-\mu_K$ are fitted to be $\SI{1.4 \pm 0.3}{ns}$, $\SI{1.1\pm 0.1}{ns}$ and $\SI{3.0\pm 0.4}{ns}$. TTS ($\SI{1.73\pm0.08}{ns}$) is defined as FWHM$=2\sqrt{2\ln 2}\sigma_{\mathrm{TT}}$~\cite{HAMAMATSUManual} representing the timing resolution. The charge and TT seem to be correlated in Fig.~\ref{fig:triggerTTS2d} and the long tail is evident in charge distribution.

\subsection{Dark count rate and pre-pulse}
\label{sec:dcr}
The dark noise mimicking PEs mainly comes from the spontaneous thermionic electrons emitted from the photocathode~\cite{KM3NetTesting}. The dark count rate (DCR) is ${N^{\mathrm{noise}}}/({N^{\mathrm{t}}T_{\mathrm{DCR}}})$, in which $N^{\mathrm{t}}$ is the number of total waveforms and $N^{\mathrm{noise}}$ is the noise count in the interval of $[\SI{-300}{ns},\SI{-150}{ns}]$ relative to the peak time $\mu_{\mathrm{TT}}$ of TT distribution with $T_{\mathrm{DCR}}=\SI{150}{ns}$. The DCR of nine MCP-PMTs is $5.8\pm\SI{1.6}{kHz}$ at room temperature.

Generated from photons hitting the MCP or the first dynode directly rather than the photocathode, \emph{pre-pulses} appear about tens of nanoseconds earlier with smaller amplitudes~\cite{JUNOMassTesting}. The probability $P_{\mathrm{pre}}$ of pre-pulses is $N^{\mathrm{pre}}/N^\mathrm{t} - \mathrm{DCR}\cdot T_{\mathrm{pre}}$, in which $N^{\mathrm{pre}}$ is the pre-pulse count of the total waveforms in the interval [\SI{-100}{ns},\SI{-10}{ns}] ($T_{\mathrm{pre}}=90$\,ns) relative to $\mu_{\mathrm{TT}}$. The small $P_{\mathrm{pre}}$ ($1\times10^{-6}\pm6\times10^{-6}$) is dominated by DCR at our low-occupancy setup.

\subsection{After-pulse}
\label{sec:afterpulse}
Ions such as \ce{H+}, \ce{He+} and \ce{O+} produced from gaseous impurities in the vacuum bulb by the PEs drift back to the photocathode, generate new electrons and then \emph{after-pulses}~\cite{JUNOMassTesting,Coates_1973}. The delay times of after-pulses are proportional to the square root of the mass-to-charge ratios of the ions~\cite{XENON1TTesting,Coates_1973,afterpulseTime}. 
\begin{figure}[!htbp]
    \centering
    \includegraphics[width=\LF\textwidth]{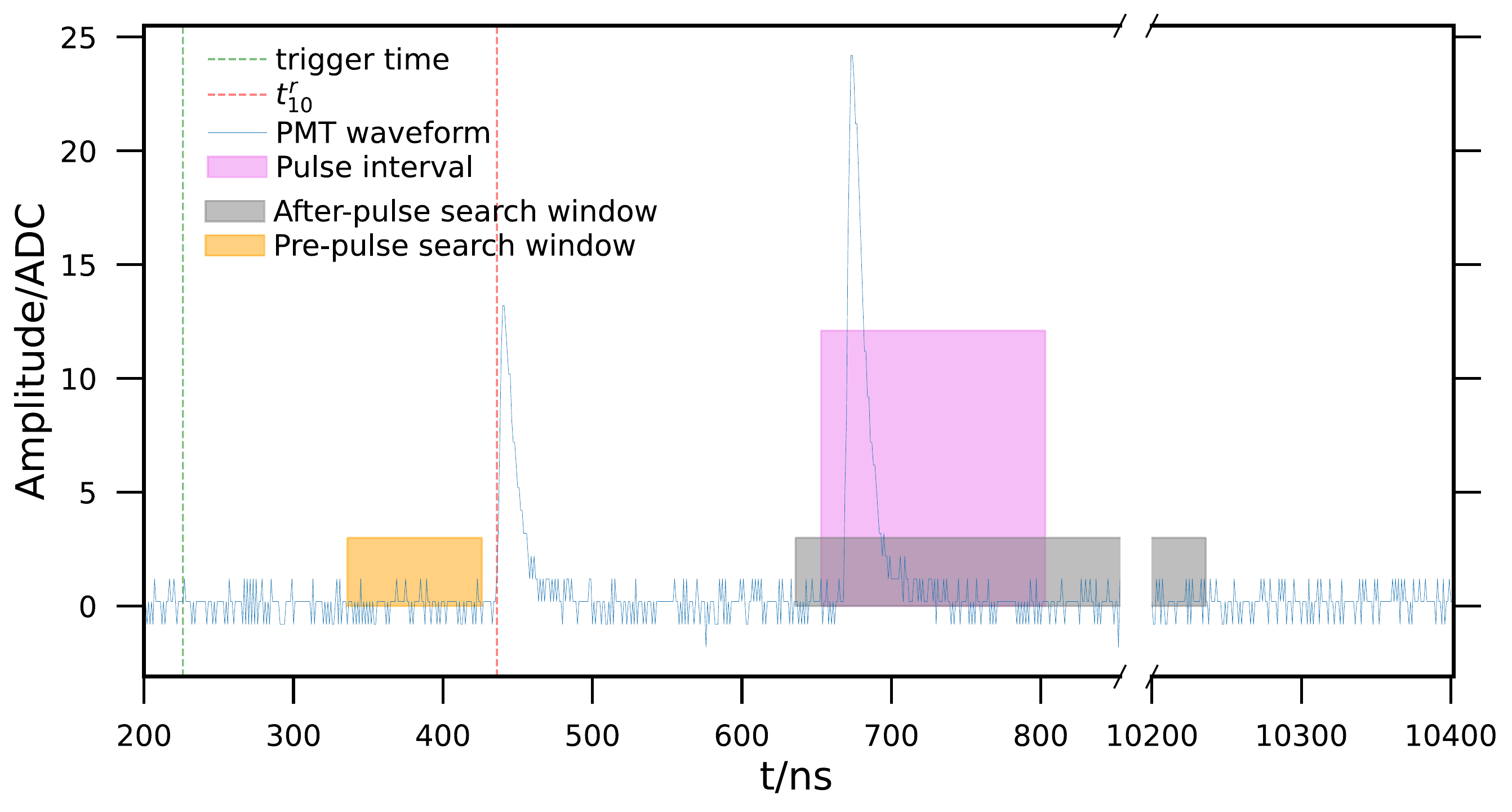}
    \caption{An example waveform for searching pre-pulses and after-pulses. Green and red vertical dashed lines are the trigger time and 10\% of the rising edge of the main pulse $t_r^{10}$, respectively. The gray and orange regions are the intervals for searching after-pulses and pre-pulses.}
    \label{fig:afterpulseSchema}
\end{figure}
The after-pulses are searched from \SI{200}{ns} after the main pulse in $N^\mathrm{hit}$ selected waveforms mentioned in Section~\ref{sec:noisepeak}. The 10\% rise time $t_r^{10}$ and charge $Q$ of the after-pulse and pre-pulse are calculated in the $[-\SI{10}{ns},\SI{75}{ns}]$ window relative to the peak position, as shown by the violet area in Fig.~\ref{fig:afterpulseSchema}.

The probability $P_{\mathrm{after}}$ of after-pulses is $N^{\mathrm{after}}/N^\mathrm{hit} - \mathrm{DCR}\cdot T_{\mathrm{after}}$, in which $N^{\mathrm{after}}$ is the pulse counts in [\SI{200}{ns}, \SI{9800}{ns}] ($T_{\mathrm{after}}=9600$\,ns) relative to the main pulses. $P_{\mathrm{after}}$ of nine MCP-PMTs is $0.009\pm0.005$.

\begin{figure}[!htbp]
    \centering
    \begin{subfigure}[t]{\LF\textwidth}
        \includegraphics[width=\textwidth]{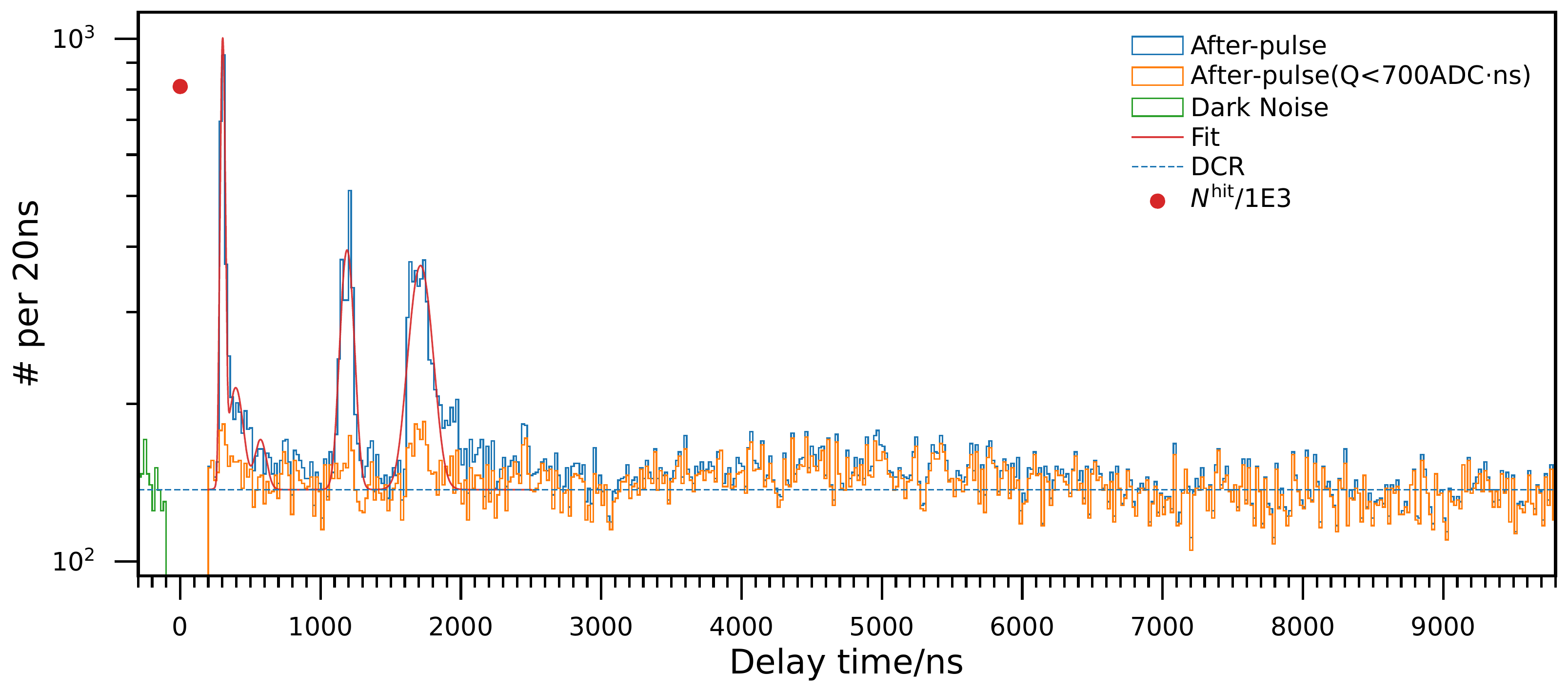}
        \caption{}
        \label{fig:afterpulse1d}
    \end{subfigure}
    \begin{subfigure}[t]{\LF\textwidth}
        \includegraphics[width=\textwidth]{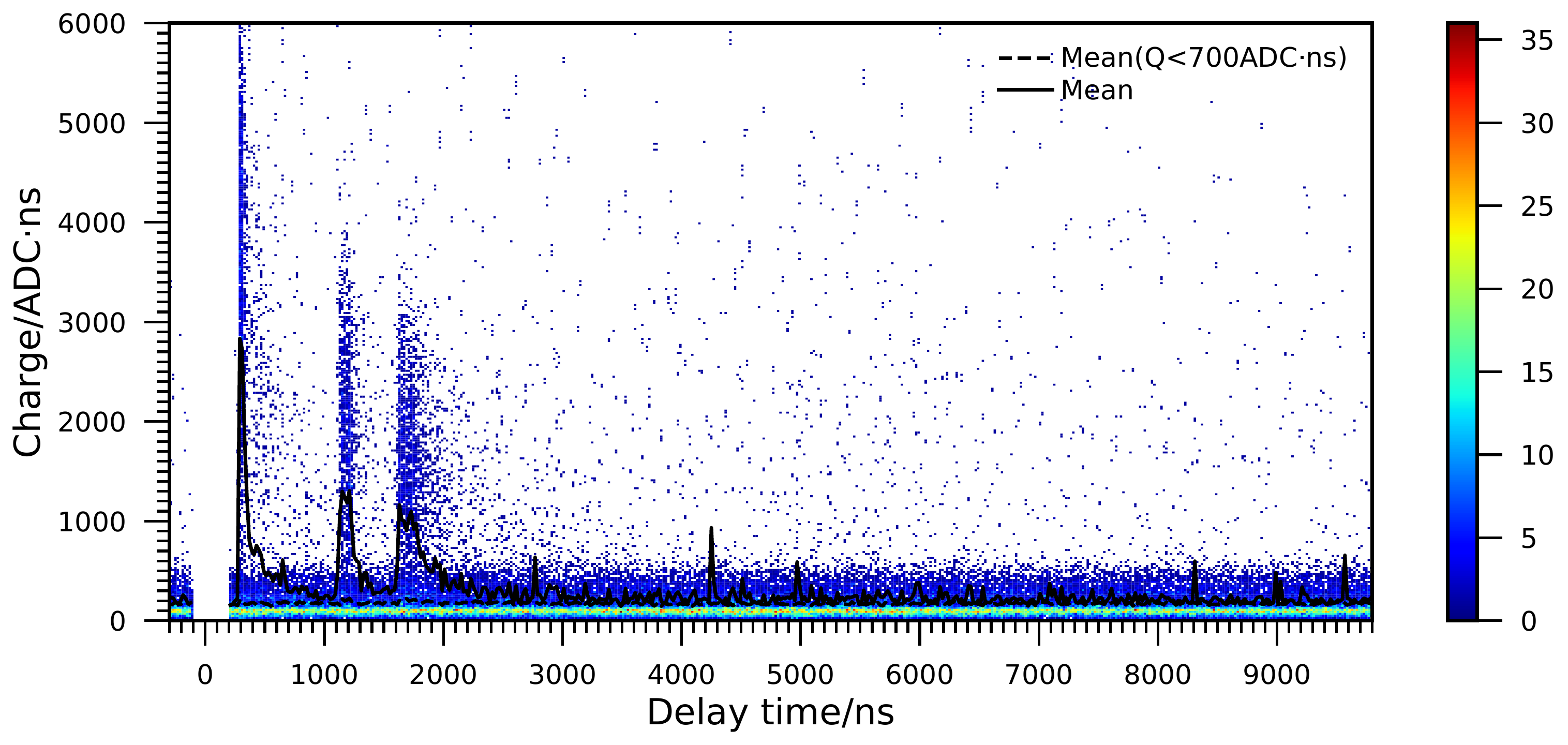}
        \caption{}
        \label{fig:afterpulse2d}
    \end{subfigure}
    \caption{(a) An example of time distributions of dark noises (green), after-pulses (blue) and after-pulses with charge cut (orange) for an MCP-PMT, where the red line shows the Gaussian fits. The red point represents $N^{\mathrm{hit}}$ at \SI{0}{ns}, around which the pre-pulses and delayed pulses are not shown. The horizontal dash line illustrates the expected dark-noise level. (b) Charge versus time distribution of pre-pulses and after-pulses. The bright horizontal band at about 100\,$\mathrm{ADC}\cdot \mathrm{ns}$ mainly contains dark noise. The black (dash) line is the mean charge of (selected) pulses in each time bin.}
\end{figure}

The distribution of the delay time from $t_r^{10}$ of the main to after-pulse in Fig.~\ref{fig:afterpulse1d} indicates five structures at around \SI{300}{ns}, \SI{400}{ns}, \SI{600}{ns}, \SI{1200}{ns} and \SI{1700}{ns}, time ratio being about $1:\sqrt{2}:\sqrt{4}:\sqrt{16}:\sqrt{32}$. These structures may originate from \ce{H^+}, \ce{H_2^+}, \ce{He^{+}}, \ce{CH_4^+}, and \ce{N_2^+} or \ce{O_2^+}. XENON1T~\cite{XENON1TTesting} and XMASS~\cite{Abe_2020} gave the same assumptions for the first peak. But similar works by JUNO~\cite{Zhao:2022gks} and KM3NeT~\cite{KM3NetTesting} assigned the first peak to \ce{H_2^{+}}.

We use five Gaussians with DCR ($\sum_{i=1}^{5}{A_if_i^{\mathrm{AP}}(t;t_i,\sigma_i^2)} + \mathrm{DCR}\cdot N^{\mathrm{hit}}$) to model the five structures, in which $A_i$, $t_i$, and $\sigma_i$ are the amplitudes, times, width of each after-pulse structure (Table.~\ref{tab:afterpulse}). 
A slow undulating structure contributes about half of the after-pulses on average in $[\SI{2000}{ns},\SI{7000}{ns}]$ as shown in Fig.~\ref{fig:afterpulse1d}, whose origin we are unable to identify.

\begin{table}
    \centering
    \caption{After-pulse parameters of MCP-PMTs}
    \label{tab:afterpulse}
    \begin{threeparttable}
        \begin{tabular}{c|ccccc}
            \hline\hline
            & \ce{H+} & \ce{H2+} & \ce{He+} & \ce{CH4+} & \ce{N2+} or \ce{O2+} \\
            \hline
            $t_i$/ns&300$\pm$4&414$\pm$25&596$\pm$53&1180$\pm$28&1714$\pm$29\\
            $A_i/N^{\mathrm{hit}}/10^{-3}$&1.6$\pm$1.1&0.5$\pm$0.3&0.2$\pm$0.2&1.1$\pm$0.7&1.9$\pm$0.6\\
            $\sigma_i$/ns&16$\pm$6&46$\pm$9 &33$\pm$10&44$\pm$11&70$\pm$20\\
            $Q_i/Q_0$&34$\pm$4&  -\tnote{*} &22$\pm$3&18$\pm$3&14$\pm$2\\
            $\sigma_{Q_i}/Q_0$&12$\pm$2&  -\tnote{*} &17$\pm$3&9$\pm$1&8$\pm$1\\
            \hline
        \end{tabular}
        \begin{tablenotes}
            \footnotesize
            \item[*] The charge of the 2nd structure cannot be fitted due to the interference with \ce{H+}.
        \end{tablenotes}
    \end{threeparttable}
\end{table}

\begin{figure}[!htbp]
    \centering
    \includegraphics[width=\textwidth]{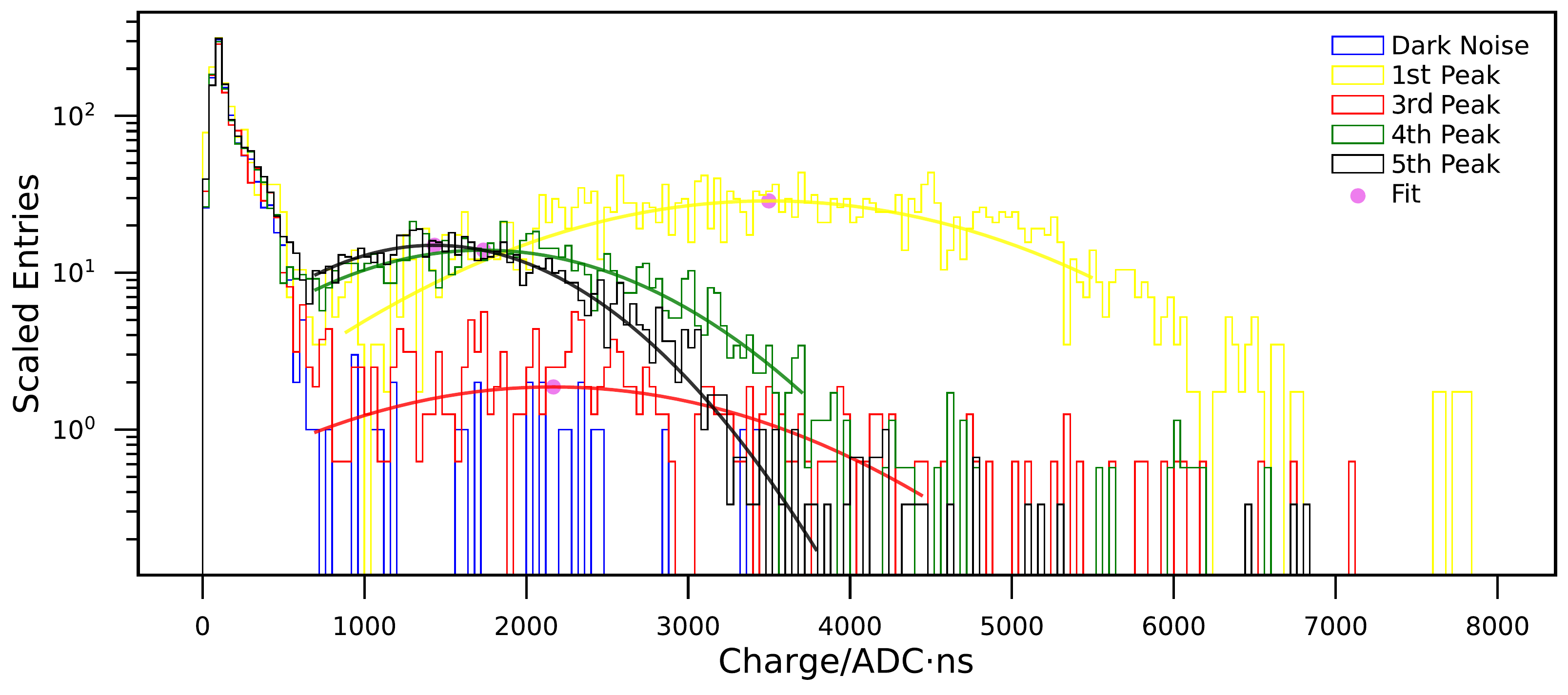}
    \caption{The charge distribution of after-pulse structures of an example MCP-PMT. The blue histogram is the distribution of dark noise. The violet points are the peaks of fit curves.}
    \label{fig:afterpulsecharge}
\end{figure}

The after-pulses contain large charge signals, as shown in Fig.~\ref{fig:afterpulse2d}. The charge distributions in $[t_i-3\sigma_i,t_i+3\sigma_i]$ of each after-pulse are scaled to the same dark noise count in Fig.~\ref{fig:afterpulsecharge}.  Charge distribution of the second structure~(\ce{H2+}) is hard to fit due to the interference with the first one~(\ce{H+}).  We fit each charge distribution with a Gaussian $f^{\rm APQ}_i(Q;Q_i,\sigma_{Q_i}^2)$, in which $Q_i$ and $\sigma_{Q_i}$ are the charge and spread of each after-pulse (Table.~\ref{tab:afterpulse}). Because the 3rd-structure charge distribution is dominated by charge less than 700\,ADC$\cdot$ns as shown in Fig.~\ref{fig:afterpulse2d} and Fig.~\ref{fig:afterpulsecharge}, $\sigma_{Q_3}$ from fitting on low-statistic entries contains large uncertainty and maybe unreliable. From the six PMTs with higher occupancies, the mean charge of each after-pulse shows a negative correlation between charge and delay time.

\subsection{Relative photon detection efficiency}
\label{sec:PDE}
A regression method is developed to combine the light-source calibration and PDE measurements simultaneously. Let $I_n$ denote the light intensity of the $n$th run, $\alpha_j$ the light allocation ratio of the $j$th splitter channel (out of four and assumed to be stable across runs), and $\epsilon_k$ the PDE of the $k$th PMT (out of one reference dynode and nine MCP-PMTs). The PE counts in each waveform follow Poisson distribution $\pi(I_n\alpha_j\epsilon_k)$.

For convenience, the index of the reference PMT is set to 0. Let $\alpha_j^0\equiv\alpha_j/{\alpha_0}$, $\epsilon_k^0\equiv\epsilon_k/{\epsilon_0}$, $I_n^0\equiv I_n\alpha_0\epsilon_0$.  The hit rate $R_{njk}$ of the $k$th PMT at the $j$th channel in the $n$th run is

\begin{equation}
    \label{equ:linkfunction}
    R_{njk}=1-\exp\left(-I_n\alpha_j\epsilon_k\right)=1-\exp\left(-e^{\log{I_n^0}+\log{\alpha_j^0}+\log{\epsilon_k^0}}\right).
\end{equation}

The number of hit waveforms $N^{\mathrm{hit}}_{njk}$ of the $k$th PMT in the $n$th run with the $j$th channel follows Binomial distribution $B(N^{\mathrm{hit}}_{njk};R_{njk},N^t_{njk})$, in which $N^t_{njk}$ is the total number of waveforms by the laser trigger. The likelihood is therefore

\begin{equation}
    \label{equ:likelihood}
    \mathcal{L}=\prod_{njk}{R_{njk}^{N^\mathrm{hit}_{njk}}(1-R_{njk})^{N^t_{njk}-N^{\mathrm{hit}}_{njk}}}.
\end{equation}

Eqs.~\eqref{equ:linkfunction} and \eqref{equ:likelihood} define a \emph{Binomial regression} with \emph{complementary log-log} link function~\cite{glm}, with $\log{\epsilon_k^0}$, $\log{\alpha_j^0}$ and $\log{I_n^0}$ as parameters. The relative PDEs $\epsilon_k^0$ of MCP-PMTs are calculated from the regression results to be about $1.71$, significantly higher than the reference PMT.  We could attribute it to the improvements on both quantum and collection efficiencies of the GDB-6082 MCP-PMTs.

\section{Energy resolution boost}
\label{Result}
Assume the PE counts $N$ on a PMT with PDE $\epsilon$ for an event with visible energy $E$ obeys Poisson distribution $\pi(\lambda_N=K\epsilon E)$, where $K$ is a factor related to light yield and detector optics. The output total charge $\sum{Q}$ is a compound Poisson random variable with the expectation $\mathrm{E}[\sum{Q}]=\lambda_N\mathrm{E}[Q]$ and variance $\mathrm{Var}[\sum{Q}]=\mathrm{E}^2[Q]\lambda_N+\lambda_N\mathrm{Var}[Q]$. The energy is estimated as $\hat{E}=\sum{Q}/(K\epsilon\mathrm{E}[Q])$ with its resolution being

\begin{equation}
    \frac{\sqrt{\mathrm{Var}[\hat{E}]}}{\mathrm{E}[\hat{E}]}=\frac{\sqrt{\mathrm{E}^2[Q]\lambda_N+\lambda_N\mathrm{Var}[Q]}}{\lambda_N\mathrm{E}[Q]}=\frac{\sqrt{1+\frac{\mathrm{Var}[Q]}{\mathrm{E}^2[Q]}}}{\sqrt{K\epsilon E}}.
\end{equation}

\begin{figure}[!htbp]
    \centering
    \includegraphics[width=\MF\textwidth]{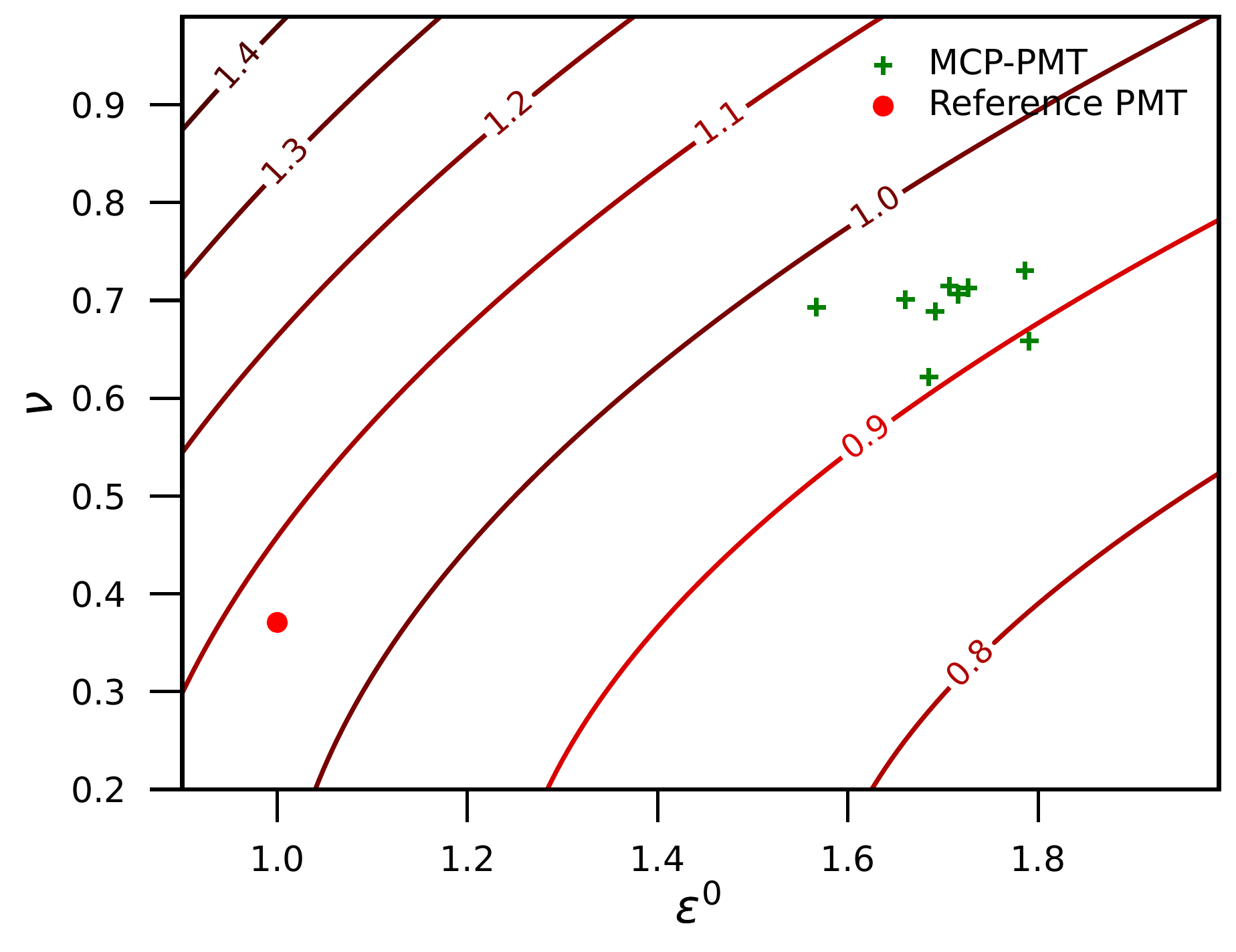}
    \caption{Contour map of energy resolution figure-of-merit $M_{E}$~(see text) as a function of the sample resolution $\nu$ and the relative PDE $\epsilon^0$. The relative PDE of the reference PMT is 1.}
    \label{fig:EnergyResolution}
\end{figure}

The PMT specific factors are $\epsilon$ (proportional to relative PDE $\epsilon^0$) and \(1 + \mathrm{Var}[Q]/ \mathrm{E}^2[Q]\) (estimated by the sample resolution $\nu^2$ in Section~\ref{sec:noisepeak}).  The latter is known as the \emph{excess noise factor}~\cite{JUNOMassTesting,ENF,ENFAuger} indicating the impact of single-PE charge smearing on energy.  We plot the \emph{figure of merit} $M_{E}\equiv\sqrt{({1+\nu^2})/{\epsilon^0}}$ of the reference and nine MCP-PMTs in Fig.~\ref{fig:EnergyResolution}. Despite the long tail in the charge distribution, the boost of PDE of MCP-PMTs leads to a 10\% better $M_{E}$.  Because of that, considering the long tail to be an undesired by-product of MCP coating to improve the collection efficiency, we decide to adopt the technology.  We are developing an advanced waveform analysis method to model the charge distribution and count PEs that may eliminate the long-tail degradation on energy resolution.

\section{Summary}
\label{Summary}
\begin{table}
    \centering
    \caption{Summary of important parameters of the MCP-PMTs}
    \label{tab:summary}
    \begin{tabularx}{\textwidth}{c|r @{$\pm$} l cXc}
        \hline\hline
        Parameters&\multicolumn{2}{c}{Value}&Criteria&Notes&Section\\
        \hline
        $\overline{Q}/Q_0$&1.8&0.1&&Entire-Sample to Main-Peak Gain Ratio&\ref{sec:noisepeak}\\
        $\nu_0=\sigma_{Q_0}/Q_0$&0.25&0.02&&Peak Resolution&\ref{sec:noisepeak}\\
        $\nu=\left.\sqrt{s^2_{Q}}\middle/\overline{Q}\right.$&0.69&0.03&&Sample Resolution&\ref{sec:noisepeak}\\
        $N^{\mathrm{1e}}/N^{\mathrm{hit}}$&0.59&0.02&&Main-peak Fraction&\ref{sec:noisepeak}\\
        P/V&5.9&1.4&$>5$&Peak-to-Valley Ratio&\ref{sec:PV}\\
        $t_r$/ns&3.71&0.15&$<4$&Rise Time&\ref{sec:SER}\\
        $t_f$/ns&15.6&1.8&$<20$&Fall Time&\ref{sec:SER}\\
        \hline
        $\sigma_{\mathrm{SER}}$/ns&1.63&0.06&& \multirow{2}{=}{Shape Parameters of SER} & \ref{sec:SER}\\
        $\tau_{\mathrm{SER}}$/ns&7.2&1.1&&&\\
        \hline
        TTS/ns&1.73&0.08&$<1.8$&Transit Time Spread &\ref{sec:TTS}\\
        DCR/kHz&5.8&1.6&$\sim 5$&Dark Count Rate&\ref{sec:dcr}\\
        $P_{\mathrm{pre}}$&1E-6&6E-6&$<0.001$&Pre-Pulse Probability&\ref{sec:afterpulse}\\
        $P_{\mathrm{after}}$&0.009&0.005&$<0.048$&After-Pulse Probability&\ref{sec:afterpulse}\\
        $\epsilon^0$&1.71&0.06&$>1.6$&Relative PDE&\ref{sec:PDE}\\
        \hline\hline
    \end{tabularx}
\end{table}

The characteristics of the nine MCP-PMTs discussed are summarized in Table~\ref{tab:summary}, which can be served as inputs to detector simulation and data analysis. A new calibration method based on regression in this study gives the average relative PDE of the MCP-PMT to be about 1.7 times the reference PMT. The long tail in the charge distribution is countered by the high PDE, resulting in an overall boost in energy resolution. We conclude that the new 8-inch GDB-6082 MCP-PMT from NNVT is suitable for the upcoming JNE.
\section{Acknowledgments}
This work is supported in part by the National Natural Science Foundation of China (12127808), the Key Laboratory of Particle \& Radiation Imaging (Tsinghua University). 

\bibliographystyle{elsarticle-num-names}
\bibliography{introduction,facility,method}

\begin{thebibliography}{47}
\expandafter\ifx\csname natexlab\endcsname\relax\def\natexlab#1{#1}\fi
\providecommand{\url}[1]{\texttt{#1}}
\providecommand{\href}[2]{#2}
\providecommand{\path}[1]{#1}
\providecommand{\DOIprefix}{doi:}
\providecommand{\ArXivprefix}{arXiv:}
\providecommand{\URLprefix}{URL: }
\providecommand{\Pubmedprefix}{pmid:}
\providecommand{\doi}[1]{\href{http://dx.doi.org/#1}{\path{#1}}}
\providecommand{\Pubmed}[1]{\href{pmid:#1}{\path{#1}}}
\providecommand{\bibinfo}[2]{#2}
\ifx\xfnm\relax \def\xfnm[#1]{\unskip,\space#1}\fi
\bibitem[{Li et~al.(2015)Li, Ji, Haxton, and Wang}]{li_second-phase_2015}
\bibinfo{author}{J.~Li}, \bibinfo{author}{X.~Ji}, \bibinfo{author}{W.~Haxton},
  \bibinfo{author}{J.~S.~Y. Wang},
\newblock \bibinfo{title}{The {Second}-phase {Development} of the {China}
  {JinPing} {Underground} {Laboratory}},
\newblock \bibinfo{journal}{Physics Procedia} \bibinfo{volume}{61}
  (\bibinfo{year}{2015}) \bibinfo{pages}{576--585}. \URLprefix
  \url{https://www.sciencedirect.com/science/article/pii/S1875389214006683}.
  \DOIprefix\doi{10.1016/j.phpro.2014.12.055}.
\bibitem[{Cheng et~al.(2017)Cheng, Kang, Li, Li, Li, Yue, Zeng, Chen, Wu, Ji,
  and Wong}]{cheng_china_2017}
\bibinfo{author}{J.-P. Cheng}, \bibinfo{author}{K.-J. Kang},
  \bibinfo{author}{J.-M. Li}, \bibinfo{author}{J.~Li}, \bibinfo{author}{Y.-J.
  Li}, \bibinfo{author}{Q.~Yue}, \bibinfo{author}{Z.~Zeng},
  \bibinfo{author}{Y.-H. Chen}, \bibinfo{author}{S.-Y. Wu},
  \bibinfo{author}{X.-D. Ji}, \bibinfo{author}{H.~T. Wong},
\newblock \bibinfo{title}{The {China} {Jinping} {Underground} {Laboratory} and
  {Its} {Early} {Science}},
\newblock \bibinfo{journal}{Annual Review of Nuclear and Particle Science}
  \bibinfo{volume}{67} (\bibinfo{year}{2017}) \bibinfo{pages}{231--251}.
  \URLprefix \url{https://doi.org/10.1146/annurev-nucl-102115-044842}.
  \DOIprefix\doi{10.1146/annurev-nucl-102115-044842}.
\bibitem[{Beacom et~al.(2017)Beacom, Chen, and et~al.}]{LetterJNE2017}
\bibinfo{author}{J.~F. Beacom}, \bibinfo{author}{S.~Chen},
  \bibinfo{author}{et~al.},
\newblock \bibinfo{title}{{Physics prospects of the {Jinping} neutrino
  experiment}},
\newblock \bibinfo{journal}{Chinese Physics C} \bibinfo{volume}{41}
  (\bibinfo{year}{2017}) \bibinfo{pages}{023002}. \URLprefix
  \url{https://doi.org/10.1088/1674-1137/41/2/023002}.
  \DOIprefix\doi{10.1088/1674-1137/41/2/023002}.
\bibitem[{Xu(2020)}]{xu_jinping_2020}
\bibinfo{author}{B.~Xu},
\newblock \bibinfo{title}{Jinping {Neutrino} {Experiment}: a {Status}
  {Report}},
\newblock \bibinfo{journal}{J. Phys.: Conf. Ser.} \bibinfo{volume}{1468}
  (\bibinfo{year}{2020}) \bibinfo{pages}{012212}. \URLprefix
  \url{https://dx.doi.org/10.1088/1742-6596/1468/1/012212}.
  \DOIprefix\doi{10.1088/1742-6596/1468/1/012212}.
\bibitem[{Xu(2022{\natexlab{a}})}]{xu_innovations_2022}
\bibinfo{author}{B.~Xu}, \bibinfo{title}{Innovations of the {Upcoming}
  {Hundred}-{Ton} {Jinping} {Neutrino} {Experiment}},
  \bibinfo{year}{2022}{\natexlab{a}}. \URLprefix
  \url{https://zenodo.org/record/6816491}.
  \DOIprefix\doi{10.5281/zenodo.6816491}.
\bibitem[{Xu(2022{\natexlab{b}})}]{xu_design_2022}
\bibinfo{author}{B.~Xu},
\newblock \bibinfo{title}{Design and {Construction} of hundred-ton liquid
  neutrino detector at {CJPL} {II}},
\newblock in: \bibinfo{booktitle}{Proceedings of 41st {International}
  {Conference} on {High} {Energy} physics — {PoS}({ICHEP2022})}, volume
  \bibinfo{volume}{414}, \bibinfo{publisher}{SISSA Medialab},
  \bibinfo{year}{2022}{\natexlab{b}}, p. \bibinfo{pages}{926}. \URLprefix
  \url{https://pos.sissa.it/414/926}. \DOIprefix\doi{10.22323/1.414.0926}.
\bibitem[{{Hamamatsu Photonics K.K.}(2017)}]{HAMAMATSUManual}
\bibinfo{author}{{Hamamatsu Photonics K.K.}}, \bibinfo{title}{{PHOTOMULTIPLIER
  TUBES: Basics and Applications FOURTH EDITION}},
  \bibinfo{howpublished}{available at
  \url{https://www.hamamatsu.com/content/dam/hamamatsu-photonics/sites/documents/99_SALES_LIBRARY/etd/PMT_handbook_v4E.pdf}
  (accessed on December 31, 2022)}, \bibinfo{year}{2017}.
\bibitem[{Bellerive et~al.(2016)}]{SNO}
\bibinfo{author}{A.~Bellerive}, et~al.,
\newblock \bibinfo{title}{{The Sudbury Neutrino Observatory}},
\newblock \bibinfo{journal}{Nuclear Physics B} \bibinfo{volume}{908}
  (\bibinfo{year}{2016}) \bibinfo{pages}{30--51}.
  \DOIprefix\doi{https://doi.org/10.1016/j.nuclphysb.2016.04.035},
  \bibinfo{note}{neutrino Oscillations: Celebrating the Nobel Prize in Physics
  2015}.
\bibitem[{Fukuda et~al.(2003)}]{SuperK}
\bibinfo{author}{S.~Fukuda}, et~al.,
\newblock \bibinfo{title}{{The Super-Kamiokande detector}},
\newblock \bibinfo{journal}{Nuclear Instruments and Methods in Physics Research
  Section A} \bibinfo{volume}{501} (\bibinfo{year}{2003})
  \bibinfo{pages}{418--462}.
  \DOIprefix\doi{https://doi.org/10.1016/S0168-9002(03)00425-X}.
\bibitem[{Gando et~al.(2011)}]{KamLAND}
\bibinfo{author}{A.~Gando}, et~al. (\bibinfo{collaboration}{The KamLAND
  Collaboration}),
\newblock \bibinfo{title}{Constraints on ${\ensuremath{\theta}}_{13}$ from a
  three-flavor oscillation analysis of reactor antineutrinos at {KamLAND}},
\newblock \bibinfo{journal}{Phys. Rev. D} \bibinfo{volume}{83}
  (\bibinfo{year}{2011}) \bibinfo{pages}{052002}. \URLprefix
  \url{https://link.aps.org/doi/10.1103/PhysRevD.83.052002}.
  \DOIprefix\doi{10.1103/PhysRevD.83.052002}.
\bibitem[{An et~al.(2016)}]{JUNO:2015zny}
\bibinfo{author}{F.~An}, et~al. (\bibinfo{collaboration}{JUNO}),
\newblock \bibinfo{title}{{Neutrino Physics with JUNO}},
\newblock \bibinfo{journal}{J. Phys. G} \bibinfo{volume}{43}
  (\bibinfo{year}{2016}) \bibinfo{pages}{030401}.
  \DOIprefix\doi{10.1088/0954-3899/43/3/030401}.
  \href{http://arxiv.org/abs/1507.05613}{{\tt arXiv:1507.05613}}.
\bibitem[{Wang et~al.(2012)}]{WANG2012113}
\bibinfo{author}{Y.~Wang}, et~al.,
\newblock \bibinfo{title}{{A new design of large area MCP-PMT for the next
  generation neutrino experiment}},
\newblock \bibinfo{journal}{Nuclear Instruments and Methods in Physics Research
  Section A: Accelerators, Spectrometers, Detectors and Associated Equipment}
  \bibinfo{volume}{695} (\bibinfo{year}{2012}) \bibinfo{pages}{113--117}.
  \URLprefix
  \url{https://www.sciencedirect.com/science/article/pii/S0168900211023199}.
  \DOIprefix\doi{https://doi.org/10.1016/j.nima.2011.12.085},
  \bibinfo{note}{new Developments in Photodetection NDIP11}.
\bibitem[{Wu et~al.(2021)}]{MCP-PMTworkgroup:2021hoy}
\bibinfo{author}{Q.~Wu}, et~al. (\bibinfo{collaboration}{MCP-PMT workgroup}),
\newblock \bibinfo{title}{{Summary of the R\&D of 20-inch MCP-PMTs for neutrino
  detection}},
\newblock \bibinfo{journal}{JINST} \bibinfo{volume}{16} (\bibinfo{year}{2021})
  \bibinfo{pages}{C11003}. \DOIprefix\doi{10.1088/1748-0221/16/11/C11003}.
\bibitem[{{Northern Night Vision Technology Ltd}(2022)}]{GDB-6082}
\bibinfo{author}{{Northern Night Vision Technology Ltd}},
  \bibinfo{title}{{Large Area MCP-PMT}}, \bibinfo{howpublished}{available at
  \url{http://www.yskjnj.com/product/common/assets/upload/2022/0302/113755f9.pdf}
  (accessed on December 31, 2022)}, \bibinfo{year}{2022}.
\bibitem[{Abusleme et~al.(2022)}]{JUNOMassTesting}
\bibinfo{author}{A.~Abusleme}, et~al. (\bibinfo{collaboration}{JUNO}),
\newblock \bibinfo{title}{{Mass testing and characterization of 20-inch PMTs
  for JUNO}},
\newblock \bibinfo{journal}{Eur. Phys. J. C} \bibinfo{volume}{82}
  (\bibinfo{year}{2022}) \bibinfo{pages}{1168}.
  \DOIprefix\doi{10.1140/epjc/s10052-022-11002-8}.
  \href{http://arxiv.org/abs/2205.08629}{{\tt arXiv:2205.08629}}.
\bibitem[{Liu(2008)}]{DayaBayTesting}
\bibinfo{author}{D.~Liu},
\newblock \bibinfo{title}{{PMT evaluation for the Daya Bay neutrino
  experiment}},
\newblock in: \bibinfo{booktitle}{2008 IEEE Nuclear Science Symposium
  Conference Record}, \bibinfo{year}{2008}, pp. \bibinfo{pages}{3133--3139}.
  \DOIprefix\doi{10.1109/NSSMIC.2008.4775017}.
\bibitem[{Calvo et~al.(2010)Calvo, Cerrada, Fernandez-Bedoya, Gil-Botella,
  Palomares, Rodriguez, Toral, and Verdugo}]{DoubleChoozeTesting}
\bibinfo{author}{E.~Calvo}, \bibinfo{author}{M.~Cerrada},
  \bibinfo{author}{C.~Fernandez-Bedoya}, \bibinfo{author}{I.~Gil-Botella},
  \bibinfo{author}{C.~Palomares}, \bibinfo{author}{I.~Rodriguez},
  \bibinfo{author}{F.~Toral}, \bibinfo{author}{A.~Verdugo},
\newblock \bibinfo{title}{{Characterization of large area photomutipliers under
  low magnetic fields: Design and performances of the magnetic shielding for
  the Double Chooz neutrino experiment}},
\newblock \bibinfo{journal}{Nucl. Instrum. Meth. A} \bibinfo{volume}{621}
  (\bibinfo{year}{2010}) \bibinfo{pages}{222--230}.
  \DOIprefix\doi{10.1016/j.nima.2010.06.009}.
  \href{http://arxiv.org/abs/0905.3246}{{\tt arXiv:0905.3246}}.
\bibitem[{Jiang et~al.(2021)}]{LHAASOTesting}
\bibinfo{author}{K.~Jiang}, et~al.,
\newblock \bibinfo{title}{{Qualification tests of 997 8-inch photomultiplier
  tubes for the water Cherenkov detector array of the LHAASO experiment}},
\newblock \bibinfo{journal}{Nucl. Instrum. Meth. A} \bibinfo{volume}{995}
  (\bibinfo{year}{2021}) \bibinfo{pages}{165108}.
  \DOIprefix\doi{10.1016/j.nima.2021.165108}.
  \href{http://arxiv.org/abs/2009.12742}{{\tt arXiv:2009.12742}}.
\bibitem[{Bronner et~al.(2020)Bronner, Nishimura, Xia, and
  Tashiro}]{HyperKTesting}
\bibinfo{author}{C.~Bronner}, \bibinfo{author}{Y.~Nishimura},
  \bibinfo{author}{J.~Xia}, \bibinfo{author}{T.~Tashiro},
\newblock \bibinfo{title}{{Development and performance of the
  20{\textquotedblright} {PMT} for Hyper-Kamiokande}},
\newblock \bibinfo{journal}{Journal of Physics: Conference Series}
  \bibinfo{volume}{1468} (\bibinfo{year}{2020}) \bibinfo{pages}{012237}.
  \URLprefix \url{https://doi.org/10.1088/1742-6596/1468/1/012237}.
  \DOIprefix\doi{10.1088/1742-6596/1468/1/012237}.
\bibitem[{Aiello et~al.(2018)}]{KM3NetTesting}
\bibinfo{author}{S.~Aiello}, et~al. (\bibinfo{collaboration}{KM3NeT}),
\newblock \bibinfo{title}{{Characterisation of the Hamamatsu photomultipliers
  for the KM3NeT Neutrino Telescope}},
\newblock \bibinfo{journal}{JINST} \bibinfo{volume}{13} (\bibinfo{year}{2018})
  \bibinfo{pages}{P05035}. \DOIprefix\doi{10.1088/1748-0221/13/05/P05035}.
\bibitem[{Barrow et~al.(2017)}]{XENON1TTesting}
\bibinfo{author}{P.~Barrow}, et~al.,
\newblock \bibinfo{title}{{Qualification Tests of the R11410-21 Photomultiplier
  Tubes for the XENON1T Detector}},
\newblock \bibinfo{journal}{JINST} \bibinfo{volume}{12} (\bibinfo{year}{2017})
  \bibinfo{pages}{P01024}. \DOIprefix\doi{10.1088/1748-0221/12/01/P01024}.
  \href{http://arxiv.org/abs/1609.01654}{{\tt arXiv:1609.01654}}.
\bibitem[{Antochi et~al.(2021)}]{XENONnTTesting}
\bibinfo{author}{V.~C. Antochi}, et~al.,
\newblock \bibinfo{title}{{Improved quality tests of R11410-21 photomultiplier
  tubes for the XENONnT experiment}},
\newblock \bibinfo{journal}{JINST} \bibinfo{volume}{16} (\bibinfo{year}{2021})
  \bibinfo{pages}{P08033}. \DOIprefix\doi{10.1088/1748-0221/16/08/P08033}.
  \href{http://arxiv.org/abs/2104.15051}{{\tt arXiv:2104.15051}}.
\bibitem[{van Eijk et~al.(2019)van Eijk, Schneider, and
  Unland}]{IceCubeTesting}
\bibinfo{author}{D.~van Eijk}, \bibinfo{author}{J.~Schneider},
  \bibinfo{author}{M.~Unland},
\newblock \bibinfo{title}{{Characterisation of Two PMT Models for the IceCube
  Upgrade mDOM}},
\newblock \bibinfo{journal}{PoS} \bibinfo{volume}{ICRC2019}
  (\bibinfo{year}{2019}) \bibinfo{pages}{1022}.
  \DOIprefix\doi{10.22323/1.358.1022}.
\bibitem[{{CAEN S.p.A.}(2022)}]{CAENV1751}
\bibinfo{author}{{CAEN S.p.A.}}, \bibinfo{title}{{V1751: 4/8 Channel 10 bit 2/1
  GS/s Digitizer}}, \bibinfo{howpublished}{available at
  \url{https://www.caen.it/products/v1751/} (accessed on December 31, 2022)},
  \bibinfo{year}{2022}.
\bibitem[{Zhang et~al.(2019)}]{JUNOPrototype}
\bibinfo{author}{H.~Q. Zhang}, et~al.,
\newblock \bibinfo{title}{{Comparison on PMT Waveform Reconstructions with JUNO
  Prototype}},
\newblock \bibinfo{journal}{JINST} \bibinfo{volume}{14} (\bibinfo{year}{2019})
  \bibinfo{pages}{T08002}. \DOIprefix\doi{10.1088/1748-0221/14/08/T08002}.
  \href{http://arxiv.org/abs/1905.03648}{{\tt arXiv:1905.03648}}.
\bibitem[{W-IE-NE-R(2022)}]{WIENERHV}
\bibinfo{author}{W-IE-NE-R}, \bibinfo{title}{{MPOD High Voltage module}},
  \bibinfo{howpublished}{available at
  \url{https://www.wiener-d.com/product/mpod-hv-module/} (accessed on December
  31, 2022)}, \bibinfo{year}{2022}.
\bibitem[{{NKT Photonics}(2022)}]{NTKLaser}
\bibinfo{author}{{NKT Photonics}}, \bibinfo{title}{{PILAS: picosecond pulsed
  diode lasers}}, \bibinfo{howpublished}{available at
  \url{https://www.nktphotonics.com/products/pulsed-diode-lasers/pilas/}
  (accessed on December 31, 2022)}, \bibinfo{year}{2022}.
\bibitem[{{CAEN S.p.A.}(2022)}]{CAENLIB}
\bibinfo{author}{{CAEN S.p.A.}}, \bibinfo{title}{{CAENDigitizer Library:
  Library of functions for CAEN Digitizers high level management}},
  \bibinfo{howpublished}{available at
  \url{https://www.caen.it/products/caendigitizer-library/} (accessed on
  December 31, 2022)}, \bibinfo{year}{2022}.
\bibitem[{{Net-SNMP Team}(2019)}]{SNMP}
\bibinfo{author}{{Net-SNMP Team}}, \bibinfo{title}{Net-{SNMP}},
  \bibinfo{howpublished}{available at \url{https://net-snmp.sourceforge.io/}
  (accessed on December 31, 2022)}, \bibinfo{year}{2019}.
\bibitem[{{PyVISA Authors}(2022)}]{VISA}
\bibinfo{author}{{PyVISA Authors}}, \bibinfo{title}{{PyVISA}: Control your
  instruments with {Python}}, \bibinfo{howpublished}{available at
  \url{https://pyvisa.readthedocs.io/en/latest/} (accessed on December 31,
  2022)}, \bibinfo{year}{2022}.
\bibitem[{{Beijing Hamamatsu Photon Techniques INC.}(2022)}]{BJBS}
\bibinfo{author}{{Beijing Hamamatsu Photon Techniques INC.}},
  \bibinfo{title}{{{CR}365-01}}, \bibinfo{howpublished}{available at
  \url{http://www.bhphoton.com/site/zh/product/guangdianqijian/duanchuangxingguangdianbeizengguan/1502326956990599170.html}
  (accessed on December 31, 2022)}, \bibinfo{year}{2022}.
\bibitem[{Kaptanoglu(2018)}]{R5912MOD}
\bibinfo{author}{T.~Kaptanoglu},
\newblock \bibinfo{title}{{Characterization of the Hamamatsu 8'' R5912-MOD
  Photomultiplier Tube}},
\newblock \bibinfo{journal}{Nucl. Instrum. Meth. A} \bibinfo{volume}{889}
  (\bibinfo{year}{2018}) \bibinfo{pages}{69--77}.
  \DOIprefix\doi{10.1016/j.nima.2018.01.086}.
  \href{http://arxiv.org/abs/1710.03334}{{\tt arXiv:1710.03334}}.
\bibitem[{Zhang et~al.(2019)}]{RCESpotlight}
\bibinfo{author}{H.~Q. Zhang}, et~al.,
\newblock \bibinfo{title}{{Study on relative collection efficiency of PMTs with
  spotlight}},
\newblock \bibinfo{journal}{Radiat Detect Technol Methods} \bibinfo{volume}{3}
  (\bibinfo{year}{2019}). \DOIprefix\doi{10.1007/s41605-019-0099-x}.
\bibitem[{Brun and Rademakers(1997)}]{brun_root_1997}
\bibinfo{author}{R.~Brun}, \bibinfo{author}{F.~Rademakers},
\newblock \bibinfo{title}{{ROOT} — {An} object oriented data analysis
  framework},
\newblock \bibinfo{journal}{Nuclear Instruments and Methods in Physics Research
  Section A: Accelerators, Spectrometers, Detectors and Associated Equipment}
  \bibinfo{volume}{389} (\bibinfo{year}{1997}) \bibinfo{pages}{81--86}.
  \URLprefix
  \url{https://www.sciencedirect.com/science/article/pii/S016890029700048X}.
  \DOIprefix\doi{10.1016/S0168-9002(97)00048-X}.
\bibitem[{Zhang et~al.(2021)}]{JUNOLongtail}
\bibinfo{author}{H.~Q. Zhang}, et~al.,
\newblock \bibinfo{title}{{Gain and charge response of 20\textquotedblright{}
  MCP and dynode PMTs}},
\newblock \bibinfo{journal}{JINST} \bibinfo{volume}{16} (\bibinfo{year}{2021})
  \bibinfo{pages}{T08009}. \DOIprefix\doi{10.1088/1748-0221/16/08/T08009}.
  \href{http://arxiv.org/abs/2103.14822}{{\tt arXiv:2103.14822}}.
\bibitem[{Cowan(1998)}]{Cowan1998StatisticalDA}
\bibinfo{author}{G.~Cowan},
\newblock \bibinfo{title}{{Statistical data analysis}},
\newblock in: \bibinfo{booktitle}{Statistical data analysis},
  \bibinfo{year}{1998}.
\bibitem[{Luo et~al.(2023)Luo, Liu, Zheng, Wang, and Chen}]{Luo:2022xrd}
\bibinfo{author}{W.~Luo}, \bibinfo{author}{Q.~Liu}, \bibinfo{author}{Y.~Zheng},
  \bibinfo{author}{Z.~Wang}, \bibinfo{author}{S.~Chen},
\newblock \bibinfo{title}{{Reconstruction algorithm for a novel Cherenkov
  scintillation detector}},
\newblock \bibinfo{journal}{JINST} \bibinfo{volume}{18} (\bibinfo{year}{2023})
  \bibinfo{pages}{P02004}. \DOIprefix\doi{10.1088/1748-0221/18/02/P02004}.
  \href{http://arxiv.org/abs/2209.13772}{{\tt arXiv:2209.13772}}.
\bibitem[{Chen et~al.(2020)}]{Furman}
\bibinfo{author}{Z.~Chen}, et~al.,
\newblock \bibinfo{title}{Analysis of secondary electron yield and energy
  spectrum of metal materials based on {Furman} model},
\newblock \bibinfo{publisher}{IEEE}, \bibinfo{year}{2020}, pp.
  \bibinfo{pages}{152--155}. \DOIprefix\doi{10.1109/ICSMD50554.2020.9261703}.
\bibitem[{Shin et~al.(2022)}]{longtail}
\bibinfo{author}{S.~Shin}, et~al.,
\newblock \bibinfo{title}{{Advances in the Large Area Picosecond Photo-Detector
  (LAPPD): 8'' x 8'' MCP-PMT with Capacitively Coupled Readout}}
  (\bibinfo{year}{2022}). \href{http://arxiv.org/abs/2212.03208}{{\tt
  arXiv:2212.03208}}.
\bibitem[{Cao et~al.(2021)}]{SecondElectron}
\bibinfo{author}{W.~Cao}, et~al.,
\newblock \bibinfo{title}{Secondary electron emission characteristics of the
  {Al2O3/MgO} double-layer structure prepared by atomic layer deposition},
\newblock \bibinfo{journal}{Ceramics International} \bibinfo{volume}{47}
  (\bibinfo{year}{2021}) \bibinfo{pages}{9866--9872}. \URLprefix
  \url{https://www.sciencedirect.com/science/article/pii/S0272884220337299}.
  \DOIprefix\doi{https://doi.org/10.1016/j.ceramint.2020.12.128}.
\bibitem[{Coates(1973)}]{Coates_1973}
\bibinfo{author}{P.~B. Coates},
\newblock \bibinfo{title}{{The origins of afterpulses in photomultipliers}},
\newblock \bibinfo{journal}{Journal of Physics D: Applied Physics}
  \bibinfo{volume}{6} (\bibinfo{year}{1973}) \bibinfo{pages}{1159--1166}.
  \URLprefix \url{https://doi.org/10.1088/0022-3727/6/10/301}.
  \DOIprefix\doi{10.1088/0022-3727/6/10/301}.
\bibitem[{Ma et~al.(2011)}]{afterpulseTime}
\bibinfo{author}{K.~J. Ma}, et~al.,
\newblock \bibinfo{title}{{Time and Amplitude of Afterpulse Measured with a
  Large Size Photomultiplier Tube}},
\newblock \bibinfo{journal}{Nucl. Instrum. Meth. A} \bibinfo{volume}{629}
  (\bibinfo{year}{2011}) \bibinfo{pages}{93--100}.
  \DOIprefix\doi{10.1016/j.nima.2010.11.095}.
  \href{http://arxiv.org/abs/0911.5336}{{\tt arXiv:0911.5336}}.
\bibitem[{Abe et~al.(2020)}]{Abe_2020}
\bibinfo{author}{K.~Abe}, et~al. (\bibinfo{collaboration}{XMASS}),
\newblock \bibinfo{title}{{Development of low-background photomultiplier tubes
  for liquid xenon detectors}},
\newblock \bibinfo{journal}{JINST} \bibinfo{volume}{15} (\bibinfo{year}{2020})
  \bibinfo{pages}{P09027}. \DOIprefix\doi{10.1088/1748-0221/15/09/P09027}.
  \href{http://arxiv.org/abs/2006.00922}{{\tt arXiv:2006.00922}}.
\bibitem[{Zhao et~al.(2022)}]{Zhao:2022gks}
\bibinfo{author}{R.~Zhao}, et~al.,
\newblock \bibinfo{title}{{Afterpulse measurement of JUNO 20-inch PMTs}}
  (\bibinfo{year}{2022}). \href{http://arxiv.org/abs/2207.04995}{{\tt
  arXiv:2207.04995}}.
\bibitem[{Hardin and Hilbe(2018)}]{glm}
\bibinfo{author}{J.~W. Hardin}, \bibinfo{author}{J.~M. Hilbe},
  \bibinfo{title}{Generalized Linear Models and Extension},
  \bibinfo{publisher}{Stata Press}, \bibinfo{year}{2018}.
\bibitem[{Teich et~al.(1986)Teich, Matsuo, and Saleh}]{ENF}
\bibinfo{author}{M.~Teich}, \bibinfo{author}{K.~Matsuo},
  \bibinfo{author}{B.~Saleh},
\newblock \bibinfo{title}{{Excess Noise Factors for Conventional and
  Superlattice Avalanche Photodiodes and Photomultiplier Tubes}},
\newblock \bibinfo{journal}{IEEE Journal of Quantum Electronics}
  \bibinfo{volume}{22} (\bibinfo{year}{1986}) \bibinfo{pages}{1184--1193}.
  \DOIprefix\doi{10.1109/JQE.1986.1073137}.
\bibitem[{Barnhill et~al.(2008)Barnhill, Suarez, Arisaka, Garcia, Gongora,
  Lucero, Navarro, Ohnuki, Risi, and Tripathi}]{ENFAuger}
\bibinfo{author}{D.~Barnhill}, \bibinfo{author}{F.~Suarez},
  \bibinfo{author}{K.~Arisaka}, \bibinfo{author}{B.~Garcia},
  \bibinfo{author}{J.~P. Gongora}, \bibinfo{author}{A.~Lucero},
  \bibinfo{author}{I.~Navarro}, \bibinfo{author}{T.~Ohnuki},
  \bibinfo{author}{A.~Risi}, \bibinfo{author}{A.~Tripathi},
\newblock \bibinfo{title}{{Testing of photomultiplier tubes for use in the
  surface detector of the Pierre Auger Observatory}},
\newblock \bibinfo{journal}{Nucl. Instrum. Meth. A} \bibinfo{volume}{591}
  (\bibinfo{year}{2008}) \bibinfo{pages}{453--466}.
  \DOIprefix\doi{10.1016/j.nima.2008.01.088}.

\end{thebibliography}
\end{document}